\documentclass[twocolumn,showkeys,showpacs,preprintnumbers,prd,superscriptaddress,nofootinbib]{revtex4-2}
\usepackage{graphicx}
\usepackage{array}
\usepackage{epsf}
\usepackage{bm}
\usepackage{amsmath}
\usepackage{amsfonts}
\usepackage{amssymb}
\usepackage{epstopdf}
\usepackage{natbib}
\usepackage{hyperref}
\usepackage{color}
\usepackage{verbatim}
\usepackage{multirow}
\usepackage{bm}
\usepackage{hyperref}
\usepackage{float}
\usepackage{tabularx}
\usepackage{booktabs}
\usepackage{threeparttable}
\usepackage[normalem]{ulem}
\usepackage{enumitem}
\usepackage{placeins}
\usepackage{booktabs}
\usepackage{adjustbox}
\usepackage{makecell}

\usepackage{xcolor}
\definecolor{darkblue}{RGB}{0,0,139} % azul escuro

\usepackage{hyperref}
\hypersetup{
    colorlinks=true, % false: boxed links; true: colored links
    linkcolor=darkblue,
    citecolor=darkblue,
    urlcolor=darkblue}
    
\makeatletter\let\expandableinput\@@input\makeatother

\newcolumntype{Y}{>{\centering\arraybackslash}X}
\renewcommand{\arraystretch}{1.75}

\begin{document}

\title{Are Cosmological Data Excluding Sterile Neutrinos or Only the Fully Thermalized Limit?}

\author{Artur Ladeira}
\email{artur.ladeira@ufrgs.br}
\affiliation{Instituto de F\'{i}sica, Universidade Federal do Rio Grande do Sul, 91501-970 Porto Alegre RS, Brazil}

\author{Rafael C. Nunes}
\email{rafadcnunes@gmail.com}
\affiliation{Instituto de F\'{i}sica, Universidade Federal do Rio Grande do Sul, 91501-970 Porto Alegre RS, Brazil}
\affiliation{Divisão de Astrofísica, Instituto Nacional de Pesquisas Espaciais, Avenida dos Astronautas 1758, São José dos Campos, 12227-010, São Paulo, Brazil}

\author{Eleonora Di Valentino}
\email{e.divalentino@sheffield.ac.uk}
\affiliation{School of Mathematical and Physical Sciences, University of Sheffield, Hounsfield Road, Sheffield S3 7RH, United Kingdom}

\author{Stefano Gariazzo}
\email{stefano.gariazzo@unito.it}
\affiliation{Instituto de F\'{i}sica Corpuscular (IFIC), CSIC‐Universitat de València, Spain}
\affiliation{University of Turin, Physics department and INFN, Sezione di Torino, Via P. Giuria 1, I–10125 Torino, Italy}

\begin{abstract}

We present a cosmological reassessment of light sterile-neutrino scenarios, examining whether current observations exclude sterile neutrinos as a class or primarily constrain the fully thermalized case. We consider three distinct realizations: (i) a fully thermalized sterile species (FTS), (ii) a different-temperature sterile-neutrino thermal relic (DTS) relative to the active neutrino background and (iii) a Dodelson--Widrow-like (DW) sterile neutrino with reduced phase-space normalization. Constraints are derived within both $\Lambda$CDM and the CPL dynamical dark-energy framework using combinations of Planck-CMB data, DESI DR2 BAO measurements, and the PantheonPlus and Union3 Type Ia supernova samples. For baseline data combinations without a local $H_0$ prior, the FTS scenario is strongly disfavored in both cosmological models. Adding the local $H_0^{\rm DN}$ prior allows $\Lambda$CDM+FTS to accommodate the high local $H_0$ value and become statistically competitive with standard $\Lambda$CDM once SNIa data are included, although the sterile-neutrino mass remains consistent with zero. By contrast, partially populated sterile-neutrino scenarios remain viable: the DW realization is broadly compatible with current observations, while the DTS scenario yields the least cosmological pressure among the cases considered. Overall, cosmological data mainly require a strongly suppressed effective sterile abundance, leading to tight constraints on \textbf{$m_s^{\rm eff}$} while allowing substantially weaker bounds on the physical sterile mass. We conclude that current observations do not generically exclude sterile neutrinos, but rather place strong pressure on fully thermalized or highly populated scenarios, highlighting the importance of production history and phase-space distribution when interpreting cosmological constraints.

\end{abstract}

%-----------------------------------------------------
%\pacs{}
%------------------------------------------------------
\maketitle
%-----------------------------------------------------

\section{Introduction}

The standard cosmological model, $\Lambda$ Cold Dark Matter ($\Lambda$CDM), provides a remarkably successful description of the Universe across a wide range of scales and epochs. Within this framework, the late-time accelerated expansion is attributed to a cosmological constant, while the matter sector is dominated by cold dark matter. Despite its impressive phenomenological success, $\Lambda$CDM leaves several fundamental questions unanswered, including the physical nature of dark matter, the origin of cosmic acceleration, and the complete structure of the neutrino sector. In this context, neutrinos remain among the most promising messengers of physics beyond the Standard Model, since their tiny but nonzero masses already require new ingredients at the particle-physics level.

From the cosmological point of view, neutrinos play a particularly important role because they affect both the background evolution and the growth of cosmic structures (see~\cite{Lesgourgues:2014zoa,Dolgov:2002wy,Abazajian:2021zui} for reviews). While still relativistic, they contribute to the radiation energy density and therefore modify the expansion rate before recombination, altering quantities such as the redshift of matter-radiation equality, the sound horizon, and the detailed structure of the CMB anisotropy spectra. Once they become non-relativistic, their large thermal velocities suppress the growth of matter perturbations below the free-streaming scale, leaving a characteristic scale-dependent imprint on the matter power spectrum and related late-time observables. For this reason, cosmological data provide one of the most sensitive probes of neutrino properties, including the absolute mass scale and the possible existence of additional light relics beyond the three active species~\cite{Hagstotz:2020ukm,DiValentino:2024xsv,Escudero:2024uea,Sabogal:2025qhz,Jiang:2024viw,DiValentino:2021hoh,Ivanov:2026dvl,Chudaykin:2025lww,Noriega:2025ulc,Poudou:2025qcx,Tanseri:2022zfe,Kumar:2022vee,RoyChoudhury:2025dhe,RoyChoudhury:2024wri,Li:2026ldf}.

Complementary constraints also come from terrestrial neutrino experiments, where light sterile neutrinos remain strongly constrained but not entirely excluded. Direct kinematic searches such as KATRIN have tightened the upper limits on the effective electron-neutrino mass and found no sterile signature in tritium beta decay, strengthening bounds on active–sterile mixing in the eV-scale regime~\cite{KATRIN:2024cdt,KATRIN:2025lph}. Likewise, accelerator experiments have further restricted this scenario: MicroBooNE reports no evidence for the single-light-sterile explanation of the LSND and MiniBooNE anomalies, while the Short-Baseline Neutrino Program and upcoming DUNE analyses will continue probing sterile-induced appearance and disappearance channels~\cite{MicroBooNE:2025nll,ICARUS:2026lvh,DUNE:2025sjq}. Together, these results increasingly challenge the simplest 3+1 interpretations, while preserving the importance of cosmological tests of sterile relics with alternative thermal histories and abundances.

Among the possible extensions of the neutrino sector, light sterile neutrinos occupy a distinguished position (see~\cite{Abazajian:2017tcc,Acero:2022wqg,Gariazzo:2015rra,Dasgupta:2021ies,Archidiacono:2022ich} for reviews of sterile neutrinos in cosmology). They arise naturally in several particle-physics constructions and have long been discussed in connection with short-baseline anomalies and other laboratory-motivated scenarios~\cite{Boyarsky:2009ix,Abazajian:2017tcc,Gariazzo:2015rra,Hagstotz:2020ukm}. In cosmology, however, the viability of sterile neutrinos depends not only on their physical mass, but also on how efficiently they are produced in the early Universe. In the simplest fully thermalized scenario, the sterile state behaves as an additional thermal relic and can contribute significantly both to the relativistic energy density at early times and to the hot dark matter budget at late times. Such a scenario is tightly constrained by cosmological observations~\cite{Feng:2025mlo,Du:2025iow,Hagstotz:2020ukm,Hu:2025lrl,DiValentino:2021rjj,Feng:2021ipq}, and analyses ranging from CMB data to the latest DESI BAO studies have generally found no robust evidence for a fully thermalized massive sterile neutrino, often reporting statistical disfavouring relative to active-neutrino extensions of comparable complexity~\cite{Ladeira:2026jne,Feng:2017nss,Pan:2023frx,Du:2025iow,Gelmini:2020ekg}.

However, this should not be naively promoted to a generic exclusion of sterile neutrinos as a class. A crucial point is that the cosmological impact of a sterile species depends on its phase-space distribution function, which is determined by its production history. Different production mechanisms can lead to distinct relations between the physical mass, the relic abundance, and the amount of extra radiation, thereby changing the cosmological signatures even when the particle content is nominally similar. It has long been recognized that, to a very good approximation, the cosmological effects of hot or warm neutrino-like relics can be characterized by a small set of effective parameters, most notably the contribution to the relativistic energy density, $\Delta N_{\rm eff}$, and an effective mass parameter controlling the late-time matter contribution~\cite{Cuoco:2005qr,Acero:2008rh,Das:2021pof}. Consequently, a claim of exclusion based on the fully thermalized limit does not automatically extend to partially thermalized or non-thermal sterile-neutrino scenarios.

Two particularly well-motivated alternatives are of special interest in this context. The first is a thermal sterile relic with a Fermi-Dirac distribution at a temperature different from that of the Standard Model neutrinos, $T_s \neq T_\nu$, as naturally realized in mirror or hidden-sector constructions in which the sterile sector has its own thermal history and need not share the temperature of the visible sector~\cite{Foot:1995pa,Berezhiani:1995yi,Feng:2008ya,Das:2010ts,Berlin:2016vnh,Bringmann:2020mgx,Nath:2024mgr}. The second is the Dodelson-Widrow (DW) scenario, in which the sterile-neutrino distribution retains the thermal shape but with a suppressed normalization controlled by the production efficiency~\cite{Dodelson:1993je,Acero:2008rh,Das:2021pof}. The latter scenario is realized through production by active-sterile neutrino oscillations, see, for example,~\cite{Gariazzo:2019gyi}. These realizations interpolate between the fully thermalized limit and highly suppressed populations, and therefore provide an ideal testing ground for addressing a sharper and more physically meaningful question: \textit{are cosmological data excluding sterile neutrinos in general, or are they predominantly excluding the assumption of full thermalization?}

Recently, in~\cite{Ladeira:2026jne}, some of us found that a massive, fully thermalized sterile component is statistically disfavored relative to the corresponding active-neutrino scenario. However, this conclusion does not, by itself, establish an equally strong case against partially thermalized or non-thermal sterile-neutrino models, whose cosmological signatures can differ substantially due to the different relationship between physical mass, abundance, and relativistic energy density~\cite{Cuoco:2005qr,Acero:2008rh,Das:2021pof}.

Motivated by this, in the present work we perform a systematic comparison of fully thermalized, thermal ($T_s \neq T_\nu$), and DW-like sterile-neutrino scenarios within both $\Lambda$CDM and dynamical dark-energy cosmologies. By confronting these realizations with current cosmological observations and analyzing their corresponding parameter constraints and model-comparison statistics, we aim to determine whether cosmological data disfavor sterile neutrinos in general or primarily the fully thermalized limit. An additional layer of model dependence arises from the cosmological background itself. Constraints on neutrino masses and extra relativistic degrees of freedom are well known to depend sensitively on the underlying cosmological model, particularly on assumptions about the dark energy sector (see, e.g.,~\cite{Yang:2017amu,Yang:2020ope,DiValentino:2021rjj,Du:2024pai}).

This paper is organized as follows. In Sec.\ref{sec:theory}, we review the theoretical framework for the sterile-neutrino scenarios considered in this work, including the fully thermalized case, the thermal relic with $T_s \neq T_\nu$, and the Dodelson-Widrow-like distribution. In Sec.\ref{sec:data}, we describe the cosmological models, datasets, and statistical methodology employed in the analysis. Our main results are presented and discussed in Sec.\ref{results}. Finally, in Sec.\ref{final}, we summarize our conclusions and discuss the broader implications for cosmological and laboratory searches for sterile neutrinos.

\section{Theoretical framework for sterile-neutrino cosmologies}
\label{sec:theory}

In this section, we review the theoretical framework adopted to describe the sterile-neutrino scenarios analyzed in this work. Our goal is to compare three physically distinct realizations of a light massive sterile species:

\begin{enumerate}[label=(\roman*)]
\item the fully thermalized limit;
\item a thermal relic with a temperature different from that of the active neutrinos; and
\item a Dodelson–Widrow-like (DW-like) scenario, characterized by a suppressed normalization of the phase-space distribution.
\end{enumerate}

These cases share the same underlying cosmological framework, namely the presence of an additional light species contributing to both the early-time radiation density and the late-time hot dark matter component. However, they differ in the mapping between physical mass, abundance, and relativistic energy density.

\subsection{General considerations}
\label{subsec:general_framework}

In standard cosmology, the total radiation density is written as
\begin{equation}
\rho_r = \rho_\gamma \left[1 + \frac{7}{8}\left(\frac{4}{11}\right)^{4/3} N_{\rm eff}\right],
\end{equation}
where $\rho_\gamma$ is the photon energy density and $N_{\rm eff}$ denotes the effective number of relativistic species. In the Standard Model, after accounting for non-instantaneous decoupling and finite-temperature effects, one expects $N_{\rm eff}^{\rm SM} \simeq 3.044$~\cite{Akita:2020szl,Froustey:2020mcq,Bennett:2020zkv}. Any departure from this reference value can be parameterized as
\begin{equation}
\Delta N_{\rm eff} \equiv N_{\rm eff} - N_{\rm eff}^{\rm SM}.
\end{equation}

For a generic sterile species with distribution function $f_s(p)$, the contribution to the early-time relativistic energy density can be expressed as~\cite{Acero:2008rh,Cuoco:2005qr}
\begin{equation}
\Delta N_{\rm eff}
\left[
\frac{7}{8}\frac{\pi^2}{15}T_\nu^4
\right]^{-1}
\frac{1}{\pi^2}\int dp\, p^3 f_s(p),
\label{eq:dneff_general}
\end{equation}
where $T_\nu$ is the active-neutrino temperature.

Here $p \equiv |\mathbf{p}|$ is the magnitude of the physical three-momentum measured in the local comoving frame, while $T_\nu$ denotes the active-neutrino temperature at the same epoch. In an expanding FRW background, $p$ redshifts as $p=q/a$, where $q$ is the corresponding comoving momentum.

At late times, once the sterile species becomes non-relativistic, its present-day physical density is conveniently written as
\begin{equation}
\omega_s \equiv \Omega_s h^2
\frac{m_s^{\rm eff}}{94.05\,{\rm eV}}
\frac{h^2 m_s}{\pi^2 \rho_c}\int dp\, p^2 f_s(p),
\label{eq:omega_general}
\end{equation}
where $m_s$ is the physical sterile-neutrino mass, $\rho_c$ is the critical density today, and $m_s^{\rm eff}$ is the effective mass parameter probed by cosmology~\cite{Acero:2008rh,Das:2021pof}.

The usefulness of the pair ${\Delta N_{\rm eff},m_s^{\rm eff}}$ lies in the fact that, to an excellent approximation at current observational precision, the cosmological impact of a hot or warm neutrino-like relic is largely determined by these two quantities. Indeed, different microscopic production mechanisms may lead to distinct momentum distributions while still generating nearly identical cosmological observables if they share the same $\Delta N_{\rm eff}$ and $m_s^{\rm eff}$~\cite{Cuoco:2005qr,Acero:2008rh,Das:2021pof}. This point is central to the logic of the present work: cosmology does not constrain merely ``a sterile neutrino’’ in the abstract, but rather specific realizations of its abundance and momentum distribution.

From a physical perspective, the role of the sterile species is twofold. At early times, increasing $\Delta N_{\rm eff}$ raises the expansion rate in the radiation era, modifying matter-radiation equality, the early Integrated Sachs-Wolfe (ISW) effect, the sound horizon, the damping scale, and the phase structure of the acoustic peaks. At late times, a massive sterile component contributes to the matter budget while preserving large thermal velocities for an extended period, thereby suppressing the growth of structures below its free-streaming scale. The relative importance of these two effects depends on how the sterile relic is populated in the early Universe and, correspondingly, on the detailed relation between $m_s$, $m_s^{\rm eff}$, and $\Delta N_{\rm eff}$.

In the following subsections, we specialize this general framework to the three sterile-neutrino realizations considered in this paper.

\subsection{Fully thermalized sterile neutrino}
\label{subsec:fully_thermalized}

We begin with the simplest and most widely used benchmark: a fully thermalized sterile neutrino. In this scenario, the sterile state is assumed to have reached thermal equilibrium in the early Universe and to follow a standard Fermi–Dirac distribution with the same temperature as the active neutrinos,
\begin{equation}
f_s(p)=\frac{1}{e^{p/T_\nu}+1}.
\label{eq:fd_standard}
\end{equation}
This is the limiting case in which the sterile population is maximal for a given thermal history and therefore typically produces the strongest cosmological signatures.

Because the sterile species has the same thermal distribution as an active neutrino, its contribution to the relativistic energy density corresponds approximately to one additional neutrino species. Therefore,
\begin{equation}
\Delta N_{\rm eff} \simeq 1,
\end{equation}
so that
\begin{equation}
N_{\rm eff} \simeq N_{\rm eff}^{\rm SM}+1 \simeq 4.044.
\end{equation}
In this limit, the late-time physical density of the sterile neutrino takes the standard thermal-relic form
\begin{equation}
\omega_s = \frac{m_s}{94.05\,{\rm eV}},
\label{eq:omega_fully_thermal}
\end{equation}
which implies
\begin{equation}
m_s^{\rm eff}=m_s.
\label{eq:meff_fully_thermal}
\end{equation}

This case is especially important because it provides the cleanest target hypothesis. It is also the most strongly constrained realization, since the sterile species simultaneously contributes maximally to the radiation density at early times and to the hot dark matter sector at late times. In practice, this means that a fully thermalized massive sterile neutrino tends to induce correlated shifts in several observables: it increases the early expansion rate, modifies matter-radiation equality and the early ISW effect, reduces the sound horizon, changes the detailed structure of the CMB damping tail, and suppresses the late-time matter power spectrum through free streaming~\cite{Abazajian:2017tcc,Boyarsky:2009ix,Escudero:2024uea}.

For completeness, the exact contribution of massive neutrinos to the background expansion can be written as
\begin{equation}
\rho_\nu(a)=\frac{T_{\nu0}^4}{2\pi^2 a^4}\sum_i \int_0^\infty dq\, q^2 \frac{\sqrt{q^2+\left(a m_{\nu i}/T_{\nu0}\right)^2}}{e^q+1}, 
\label{eq:rho_nu_exact} 
\end{equation}

which smoothly interpolates between the relativistic regime, $\rho_\nu \propto a^{-4}$, and the non-relativistic regime, $\rho_\nu \propto a^{-3}$. In the present context, the sterile species simply adds one extra term of the same form, with mass $m_s$, to the total neutrino sector. This is precisely why the fully thermalized case can be viewed as the most direct extension of the standard cosmological neutrino background.

From a physical standpoint, the fully thermalized limit is also the most restrictive one for a very simple reason: it maximizes both the number density and the early relativistic energy density of the sterile species. Consequently, if current cosmological data already disfavour this limit, that conclusion should be interpreted specifically as evidence against the assumption of \emph{full thermalization}, and not immediately as a universal exclusion of all sterile-neutrino realizations. This distinction motivates the alternative scenarios considered below.

\subsection{Thermal sterile relic with a different decoupling temperature}
\label{subsec:thermal_different_temperature}

A natural generalization of the previous case is obtained when the sterile neutrino has a thermal Fermi–Dirac distribution, but with a temperature $T_s$ different from that of the standard active-neutrino sector. Such a situation can arise, for instance, if the sterile state belongs to a partially decoupled or hidden sector with its own thermal history~\cite{Feng:2008ya,Das:2010ts,Acero:2008rh,Das:2021pof}. In that case, the phase-space distribution is also given by Eq.(\ref{eq:fd_standard}), but its contribution to the effective number of relativistic species is~\cite{Acero:2008rh,Das:2021pof}

\begin{equation}
\Delta N_{\rm eff}=\left(\frac{T_s}{T_\nu}\right)^4,
\label{eq:dneff_Ts}
\end{equation}
where $T_\nu$ denotes the standard neutrino temperature in the instantaneous-decoupling limit. On the other hand, once the species becomes non-relativistic, its number density scales as $T_s^3$, so that the present-day physical density is~\cite{Acero:2008rh,Das:2021pof}
\begin{equation}
\omega_s=\frac{m_s}{94.05\,{\rm eV}}
\left(\frac{T_s}{T_\nu}\right)^3.
\label{eq:omega_Ts}
\end{equation}
Combining these two expressions leads to the usual relation~\cite{Acero:2008rh,Das:2021pof}
\begin{equation}
m_s^{\rm eff}=m_s\,(\Delta N_{\rm eff})^{3/4}.
\label{eq:meff_Ts}
\end{equation}

These relations show immediately why this scenario is phenomenologically distinct from the fully thermalized case. If $T_s<T_\nu$, then the sterile population is colder and less abundant. As a result, the same physical mass $m_s$ produces a smaller contribution to the early radiation density and a smaller late-time matter density than in the fully thermalized limit. Equivalently, for fixed $m_s^{\rm eff}$, the physical mass $m_s$ can be substantially larger than the fully thermalized value. This weakens the naive one-to-one correspondence between an eV-scale sterile mass and its cosmological impact~\cite{Acero:2008rh,Das:2021pof}.

The physical interpretation is transparent. The factor $(T_s/T_\nu)^4$ controls the amount of radiation carried by the sterile species while it is relativistic, whereas the factor $(T_s/T_\nu)^3$ controls its number density once it becomes non-relativistic. Therefore, a colder thermal relic contributes less to $N_{\rm eff}$ than a fully thermalized species and, for the same physical mass, also contributes less to the present-day hot dark matter density. This is precisely why colder thermal sterile neutrinos can evade part of the cosmological pressure that strongly constrains the fully thermalized limit~\cite{Acero:2008rh,Das:2021pof}.

At the same time, the cosmological signatures remain qualitatively similar: the sterile relic still affects both the early expansion rate and the late-time growth of structure. What changes is the quantitative relation between the observable parameters $\Delta N_{\rm eff}$ and $m_s^{\rm eff}$ and the underlying microscopic mass $m_s$. This scenario is therefore an ideal benchmark for addressing the central question of this work: whether current data are excluding sterile neutrinos as a class, or primarily excluding the fully thermalized limit.

\subsection{Dodelson--Widrow-like sterile neutrino}
\label{subsec:dwlike}

We next consider the Dodelson–Widrow-like realization, in which the sterile neutrino is produced out of equilibrium and inherits the same momentum dependence as a thermal Fermi–Dirac distribution, but with a suppressed overall normalization~\cite{Dodelson:1993je,Acero:2008rh}. The distribution function is then
\begin{equation}
f_s(p)=\chi\,\frac{1}{e^{p/T_\nu}+1},
\label{eq:fd_dw}
\end{equation}
where $\chi$ is an effective suppression factor describing the incomplete population of the sterile state.

This parameterization is especially convenient because both the relativistic energy density and the late-time number density scale linearly with the overall normalization of the distribution. One therefore obtains
\begin{equation}
\Delta N_{\rm eff}=\chi,
\label{eq:dneff_dw}
\end{equation}
and
\begin{equation}
m_s^{\rm eff}= \chi\, m_s.
\label{eq:meff_dw}
\end{equation}
Equivalently,
\begin{equation}
\omega_s = \frac{m_s\chi}{94.05\,{\rm eV}} .
\label{eq:omega_dw}
\end{equation}

Compared to the thermal case with $T_s \neq T_\nu$, the DW-like scenario suppresses the sterile abundance through a multiplicative normalization rather than through a reduced temperature. This difference is microscopically important because it corresponds to a distinct production mechanism, but at the level of current cosmological observables the distinction is often subdominant once $\Delta N_{\rm eff}$ and $m_s^{\rm eff}$ are fixed~\cite{Acero:2008rh,Das:2021pof}. In other words, a DW-like sterile neutrino and a colder thermal sterile neutrino can be nearly degenerate from the perspective of present-day cosmological fits, provided they share the same effective parameters.

The physical meaning of Eq.~(\ref{eq:meff_dw}) is straightforward. The suppression factor $\chi$ simultaneously reduces the radiation density at early times and the late-time matter density associated with the sterile species. Therefore, even if the physical sterile mass $m_s$ is in the eV range, its cosmological impact can be much weaker than in the fully thermalized case whenever $\chi \ll 1$. This is the key reason why the phrase ``cosmology excludes sterile neutrinos’’ can be misleading if interpreted without qualification: what is often tightly constrained is the combination of large abundance and large mass implied by the fully populated thermal limit, not necessarily every partially populated realization.

The DW-like case is thus complementary to the thermal $T_s \neq T_\nu$ scenario. In the latter, the suppression arises through a colder thermal bath; in the former, it arises through an overall depletion of the distribution function. Both scenarios provide physically motivated ways of interpolating between the fully thermalized limit and highly suppressed sterile populations, and both are essential for a meaningful assessment of how general current cosmological bounds on sterile neutrinos really are.

\section{Datasets and methodology}
\label{sec:data}

This section presents the observational datasets and the statistical methodology employed to constrain the cosmological scenarios investigated in this work. Since our goal is to assess whether current cosmological data are excluding sterile neutrinos generically or primarily the fully thermalized limit, we analyze the same observational combinations across different sterile-neutrino realizations and within two cosmological backgrounds, namely $\Lambda$CDM and the Chevallier–Polarski–Linder (CPL) model~\cite{Chevallier:2000qy,Linder:2002et}. We begin by introducing the key datasets used throughout this analysis:

\begin{enumerate}

\item {\bf Cosmic Microwave Background (CMB):}
For the CMB, we employ the 2018 \textit{Planck} legacy release, including temperature and polarization angular power spectra, their cross-correlations, and the reconstructed lensing signal~\cite{Planck:2018vyg}. More specifically, we use the high-multipole \texttt{Plik} likelihood for TT in the interval $30 \leq \ell \leq 2508$, and for TE and EE over $30 \leq \ell \leq 1996$. These data are combined with the low-multipole temperature likelihood covering $2 \leq \ell \leq 29$, as well as the low-$\ell$ polarization likelihood based on \texttt{SimAll} in the same multipole range~\cite{Planck:2019nip}. In addition, we include the \textit{Planck} lensing reconstruction obtained from the four-point temperature correlation function~\cite{Planck:2018lbu}. Throughout this paper, this full CMB dataset is referred to simply as \textbf{CMB}.

\item {\bf Baryon Acoustic Oscillations (DESI-DR2):}
To constrain the late-time expansion history, we make use of baryon acoustic oscillation measurements from the second data release of DESI. These measurements include BAO information derived from galaxy and quasar tracers~\cite{DESI:2025zgx}, together with the contribution from Lyman-$\alpha$ forest observations~\cite{DESI:2025zpo}. The corresponding measurements, reported in \textbf{Table~\ref{tab:thermal_relic_cpl_lcdm}} of Ref.~\cite{DESI:2025zgx}, cover the effective redshift range $0.295 \leq z \leq 2.330$ and are organized into nine redshift bins. The observables are expressed in terms of the ratios $D_{\rm M}/r_d$, $D_{\rm H}/r_d$, and $D_{\rm V}/r_d$, where $D_{\rm M}$ is the transverse comoving distance, $D_{\rm H}$ is the Hubble distance, $D_{\rm V}$ is the volume-averaged distance, and $r_d$ denotes the sound horizon at the drag epoch. The full covariance matrix is taken into account, including the correlations among the quoted BAO quantities through the coefficients $r_{\rm V,M/H}$ and $r_{\rm M,H}$. In what follows, this dataset will be denoted as \textbf{BAO DESI DR2}.

\item {\bf Type Ia Supernovae (SNIa):}
As complementary probes of the late-time background evolution, we consider two modern SNIa compilations. The first is the PantheonPlus sample~\cite{Brout:2022vxf}, which consists of 1701 light-curve measurements corresponding to 1550 distinct supernovae in the redshift interval $0.01 \leq z \leq 2.26$; we refer to this compilation as \textbf{PP}. The second is the Union3.0 catalog\cite{Rubin:2023jdq}, containing 2087 SNIa spanning the range $0.001 < z < 2.26$, of which 1363 are shared with PantheonPlus; this sample will be denoted as \textbf{Union3}.\footnote{The Union3 dataset is constructed using a Bayesian hierarchical approach designed to consistently incorporate observational uncertainties and systematic effects.} Since these two samples are not statistically independent, PantheonPlus and Union3 are not combined in the same likelihood analysis.

\end{enumerate}

In the baseline cosmological setup, the active-neutrino sector is described by the minimal-mass configuration allowed by oscillation data, namely $\sum m_\nu=0.06~{\rm eV}$, consistent with the normal mass ordering. As commonly adopted in cosmological analyses, this minimal configuration is implemented as one massive active neutrino state carrying the total mass $m_\nu=0.06~{\rm eV}$, together with two effectively massless active states.
It has been demonstrated that even in the most optimistic scenarios considering incoming and future cosmological observations, the individual mass eigenstates cannot be distinguished within the expected sensitivity \cite{Archidiacono:2020dvx}.
As a consequence, at the cosmological level it is impossible to determine whether there are one massive and two massless, one massless and two massive neutrino states, or three (possibly degenerate) massive active neutrinos.
Since the upper limit on the sum of the neutrino masses obtained when considering DESI data together with CMB observations is very close to the value we assume as fixed (0.06~eV versus $\sum m_{\nu,\mbox{ active}}<0.072$~eV at 95\% CL~\cite{DESI:2024mwx}, see also e.g.~\cite{Jiang:2024viw}),
the range of allowed values for active neutrino masses is quite limited.
As a consequence, we do not expect a significant variation of the limits on the sterile neutrino component on different assumptions about the active neutrino sector, i.e.\ on a varying active neutrino mass sum or a different configuration for the mass eigenstates.

Relative to this fiducial active sector, we investigate the impact of extending the neutrino sector by introducing one additional sterile species under different assumptions for its phase-space distribution. Thus, in the sterile-neutrino extensions considered below, the neutrino sector consists of two effectively massless active states, one massive active state with a fixed mass of $0.06~{\rm eV}$, and one additional sterile state whose mass, abundance, and temperature depend on the specific production scenario.

At the level of the cosmological background, we consider two distinct frameworks. The first is the standard $\Lambda$CDM model. The second is the CPL parameterization, in which the dark-energy equation of state is written as
\begin{equation}
w(a)=w_0+w_a(1-a).
\end{equation}
This choice allows us to test whether the cosmological status of sterile neutrinos is robust against a phenomenologically relevant extension of the late-time dark-energy sector.

\begin{table}
	\begin{center}
		\renewcommand{\arraystretch}{1.4}
		\begin{tabular}{c@{\hspace{1 cm}}@{\hspace{1 cm}} c }
			\hline
			\textbf{Parameter}           & \textbf{Prior}\\
			\hline

			$\omega_{b}$                 & $[0.01,\,1.0]$ \\ 
			$\omega_{\rm cdm}$           & $[0.01,\,1.0]$ \\ 
			$\tau_{\rm reio}$            & $[0.004,\,0.8]$ \\ 
			$n_s$                        & $[0.1,\,2.0]$ \\ 
			$\ln(10^{10} A_s)$           & $[1.0,\,5.0]$ \\ 
			$H_0$                        & $[20,\,100]$ \\ 
			\hline

			$w_a$ \;          & $[-5,\,0]$ \\ 
			$w_0$ \;          & $[-5,\,5]$ \\ 
			\hline

			$m_s\;[\mathrm{eV}]$ \; 
			                             & $[0.0,\,5.0]$ \\ 
			$T_s$ \; (DTS only) 
			                             & $[0.05,\,0.72]$ \\ 
			$\chi$ \; (DW-like only) 
			                             & $[0.0,\,1.0]$ \\ 
			
			\hline
		\end{tabular}
	\end{center}
	\caption{Flat priors imposed on the free cosmological and model parameters used in the statistical analyses. The parameters $w_0$ and $w_a$ are varied only in the CPL cosmology. In the sterile-neutrino sector, the physical mass $m_s$ is sampled in the extended scenarios. For the DTS realization, the sterile temperature parameter $T_s$ is varied while the normalization of the sterile distribution is fixed to unity. The corresponding relativistic contribution is derived as $\Delta N_{\rm eff}=(T_s/0.716)^4$. For the DW-like realization, the suppression factor $\chi$ is varied while $T_\nu=0.716$ is kept fixed, so that $\Delta N_{\rm eff}=\chi$. In both partially populated scenarios, $m_s^{\rm eff}$ and $\Delta N_{\rm eff}$ are derived parameters.}
	\label{tab:priors}
\end{table}

Accordingly, the DTS and DW-like cases are sampled directly in terms of the parameters specifying their respective phase-space distributions ($T_s$ and $\chi$). The effective quantities $\Delta N_{\rm eff}$ and $m_s^{\rm eff}$, which are more directly connected to the cosmological observables, are derived from the sampled parameters according to the relations introduced in Sec.~\ref{sec:theory}.

The exact set of sampled cosmological parameters and their prior ranges are listed in Table~\ref{tab:priors}, which we retain in the same format as in our previous analyses. For convenience, the baseline parameter vectors can be schematically written as
\begin{equation}
\Theta_{\Lambda{\rm CDM}}=
\left\{
\omega_b,\,
\omega_{\rm cdm},\,
H_0,\,
\ln(10^{10}A_s),\,
n_s,\,
\tau_{\rm reio}
\right\},
\end{equation}
and
\begin{equation}
\Theta_{{\rm CPL}}=
\left\{
\omega_b,\,
\omega_{\rm cdm},\,
H_0,\,
\ln(10^{10}A_s),\,
n_s,\,
\tau_{\rm reio},\,
w_0,\,
w_a
\right\}.
\end{equation}
To these baseline parameter sets, we add the neutrino-sector parameters appropriate to each sterile-neutrino realization.
More specifically, the scenarios investigated in this work are the following:

\begin{itemize}

\item \textbf{Fully thermalized sterile neutrino (FTS):}
In this case, the sterile species is assumed to have the same Fermi--Dirac temperature and normalization as an active neutrino species. Therefore, its abundance is fully fixed by the assumption of complete thermalization, corresponding approximately to $\Delta N_{\rm eff}\simeq 1$, while the only free sterile-sector parameter is the physical sterile mass $m_s$. In the \texttt{CLASS}~\cite{Blas:2011rf} implementation, this corresponds to an additional fully populated massive species, while the active sector remains fixed to the minimal-mass configuration. Hereafter, the abbreviation \texttt{FTS} will be used to denote this scenario in the cosmological context.

\item \textbf{Thermal sterile relic with a different temperature (DTS):}
Here, the sterile neutrino is assumed to follow a thermal Fermi--Dirac distribution, but with a temperature $T_s$ different from that of the standard active-neutrino sector. In this case, the sampled sterile-sector parameters are the physical mass $m_s$ and the sterile temperature parameter $T_s$ (or, equivalently, the temperature ratio $T_s/T_\nu$). Hereafter, the abbreviation \texttt{DTS} will be used to denote this scenario in the cosmological context.

\item \textbf{DW-like sterile neutrino:}
Finally, we consider a Dodelson--Widrow-like realization, in which the sterile-neutrino distribution has the same thermal shape as the active one, but with a suppressed overall normalization. In this case, the sampled sterile-sector parameters are the physical mass $m_s$ and the suppression factor $\chi$, which, in the numerical implementation, is encoded through the effective phase-space normalization of the extra species. Hereafter, the abbreviation \texttt{DW} will be used to denote this scenario in the cosmological context.

\end{itemize}

Although the numerical implementation is specified in terms of the parameters defining each phase-space distribution, the quantities most directly constrained by cosmological observables are $\Delta N_{\rm eff}$ and $m_s^{\rm eff}$. As discussed in Sec.~\ref{sec:theory}, in the DTS scenario the sampled parameters $\{m_s,T_s\}$ are mapped into the effective quantities $\Delta N_{\rm eff}$ and $m_s^{\rm eff}$ through the temperature ratio of the sterile relic, whereas in the DW-like scenario the sampled parameters $\{m_s,\chi\}$ are mapped through the normalization factor $\chi$ of the active-neutrino distribution.

Accordingly, we regard $\Delta N_{\rm eff}$ and $m_s^{\rm eff}$ as the fundamental cosmological parameters. Because flat priors are imposed on the production parameters of each scenario, the induced prior distribution in the $(\Delta N_{\rm eff}, m_s^{\rm eff})$ plane depends on the scenario-dependent mappings between the sampled production parameters and the effective cosmological quantities, as defined in Sec.~\ref{sec:theory}. Therefore, the marginalized constraints and Bayesian evidences for the DTS and DW models should be interpreted as comparisons between alternative sterile-neutrino production histories, rather than as prior-independent cosmological discrimination among momentum distributions at fixed $\Delta N_{\rm eff}$ and $m_s^{\rm eff}$.

Thus, for each cosmological background ($\Lambda$CDM or CPL), we analyze the \texttt{FTS}, \texttt{DTS}, and \texttt{DW} sterile-neutrino extensions using the same observational combinations and the same statistical pipeline. The purpose of this strategy is to isolate the role played by the sterile thermal history, independently of the assumed late-time cosmological background.

In order to constrain the parameter space of the proposed scenarios, we employ the publicly available Boltzmann solver \texttt{CLASS}~\cite{Blas:2011rf} to compute the cosmological background evolution and linear perturbations for each sterile-neutrino realization and cosmological model considered. Parameter inference is carried out using the Markov Chain Monte Carlo (MCMC) framework implemented in \texttt{MontePython}~\cite{Audren:2012wb,Brinckmann:2018cvx}. The minimum $\chi^2$ values quoted in this work correspond to the maximum-likelihood points identified within the converged MCMC chains, following the default procedure adopted by \texttt{MontePython}. The convergence of the chains is assessed using the Gelman--Rubin criterion~\cite{Gelman:1992zz}, requiring
\begin{equation}
R-1 < 10^{-2}.
\end{equation}

Unless otherwise stated, all quoted uncertainties correspond to the $68\%$ CL, while upper limits on sterile-neutrino masses are reported at the $95\%$ CL.

We close this section by describing the statistical estimators adopted to quantify the observational performance of the extended models. We consider two complementary indicators: the minimum effective chi-square, $\chi^2_{\rm min}$, and the Akaike Information Criterion (AIC). The latter is defined as~\cite{Akaike:1974vps}
\begin{equation}
{\rm AIC}=-2\ln {\cal L}_{\rm max}+2n_0,
\end{equation}
where ${\cal L}_{\rm max}$ is the maximum likelihood and $n_0$ is the number of free model parameters.

For the model-comparison quantities reported below, each
sterile-neutrino extension is compared with the corresponding baseline
cosmology without the sterile component, while keeping the same
late-time background. Thus, $\Lambda$CDM+FTS,
$\Lambda$CDM+DTS, and $\Lambda$CDM+DW are compared against baseline
$\Lambda$CDM, whereas CPL+FTS, CPL+DTS, and CPL+DW are compared
against baseline CPL. For each model and dataset combination, we define
\begin{equation}
\Delta \chi^2_{\rm min}
=
\chi^2_{\rm min}({\cal M}_{\rm sterile})
-
\chi^2_{\rm min}({\cal M}_{\rm base}),
\label{eq:dchi2}
\end{equation}
and
\begin{equation}
\Delta {\rm AIC}
=
{\rm AIC}({\cal M}_{\rm sterile})
-
{\rm AIC}({\cal M}_{\rm base}).
\label{eq:daic}
\end{equation}
With this convention, negative values favor the sterile extension, whereas positive values favor the corresponding baseline cosmology. The AIC penalty explicitly accounts for the additional sterile-sector parameters introduced in each realization. In particular, the FTS case introduces one additional sampled sterile-sector parameter, $m_s$, while the DTS and DW-like cases introduce two additional sampled sterile-sector parameters, $\{m_s,T_s\}$ and $\{m_s,\chi\}$, respectively.

In addition to the indicators discussed above, we also perform a Bayesian model-comparison analysis through the computation of the Bayesian evidence. For a given model ${\cal M}$ with parameter vector $\boldsymbol{\theta}$, the evidence is defined as
\begin{equation}
{\cal Z} \equiv P({\bf D}\vert {\cal M})=
\int {\cal L}({\bf D}\vert \boldsymbol{\theta},{\cal M})\,
\pi(\boldsymbol{\theta}\vert {\cal M})\, d\boldsymbol{\theta},
\label{eq:evidence}
\end{equation}
where ${\cal L}$ is the likelihood function, $\pi(\boldsymbol{\theta}\vert {\cal M})$ is the prior distribution, and ${\bf D}$ denotes the dataset. The ratio of the evidences for two competing models, ${\cal M}_i$ and ${\cal M}_j$, defines the Bayes factor,
\begin{equation}
B_{ij} \equiv \frac{{\cal Z}_i}{{\cal Z}_j},
\qquad
\ln B_{ij} = \ln {\cal Z}_i - \ln {\cal Z}_j.
\label{eq:bayes_factor}
\end{equation}
A positive value of $\ln B_{ij}$ indicates that the data favor model ${\cal M}_i$ over the reference model ${\cal M}_j$, while a negative value indicates the opposite.

For guidance, we interpret the magnitude of the Bayes factor using a
Jeffreys-type scale~\cite{Jeffreys:1961,Kass:1995loi}: values
$|\ln B_{ij}|<1$ are regarded as inconclusive, $1\leq |\ln B_{ij}|<2.5$
as weak evidence, $2.5\leq |\ln B_{ij}|<5$ as moderate evidence, and
$|\ln B_{ij}|\geq5$ as strong evidence in favor of the model selected
by the sign of $\ln B_{ij}$.
Unlike $\chi^2_{\rm min}$-based estimators, the Bayesian evidence explicitly depends on the adopted prior volume and therefore penalizes unnecessarily large parameter spaces through marginalization over the full parameter domain.

To estimate the Bayesian evidence, we make use of the public code \texttt{MCEvidence}~\cite{Heavens:2017afc,MCEvidenceGitHub}, which implements the method proposed in Ref.~\cite{Heavens:2017afc} for extracting marginal likelihoods directly from MCMC chains. The algorithm estimates the local density of chain samples in parameter space using $k$th nearest-neighbor distances and thereby allows one to reuse standard MCMC chains, originally generated for parameter inference, to obtain the model evidence~\cite{Heavens:2017afc}. In the present work, we use \texttt{MCEvidence} to compute the Bayesian evidence for each model and dataset combination, and we report the corresponding Bayes factors with respect to the appropriate reference cosmology.

More precisely, in direct analogy with Eqs.~(\ref{eq:dchi2}) and~(\ref{eq:daic}), we evaluate
\begin{equation}
\ln B_{ij} = \ln {\cal Z}({\cal M}_i) - \ln {\cal Z}({\cal M}_j),
\label{eq:lnbij}
\end{equation}
where ${\cal M}_i$ denotes a sterile-neutrino extension of a given cosmological background, and ${\cal M}_j$ denotes the corresponding baseline model without the sterile component. Thus, CPL sterile-neutrino scenarios are compared against baseline CPL, while $\Lambda$CDM sterile-neutrino scenarios are compared against baseline $\Lambda$CDM. In this setup, the Bayes factor directly measures whether the data favor the sterile-neutrino extension within a fixed background cosmology. Since the Bayesian evidence already incorporates an Occam penalty through marginalization over the prior volume, no explicit matching in the number of free parameters is required, although the result remains sensitive to the prior choices adopted for each model. In the present analysis, the evidence calculation is performed using the same prior ranges adopted in the corresponding parameter-estimation runs.

This strategy is particularly important for the interpretation of the results presented below. Since the central question of this paper is whether cosmological data are excluding sterile neutrinos in general or only the fully thermalized limit, the relevant comparison is not simply between a sterile model and a minimal baseline, but also among different sterile-neutrino realizations. Only under this condition can one robustly assess whether the statistical pressure arises specifically from the assumption of full thermalization, or whether it persists across more general sterile-neutrino realizations.

\section{Main Results}
\label{results}

In what follows, we discuss each of the three cases introduced above in a separate subsection.

\subsection{Fully thermalized}

% ============================
% Table: Sterile for CPL and LCDM
% ============================

\begin{table*}[!t]
\centering
\caption{
Marginalized constraints and posterior mean values at the $68\%$ CL for the fully thermalized sterile-neutrino scenario in the CPL and $\Lambda$CDM cosmologies, using Planck CMB data combined with BAO DESI DR2, PantheonPlus (PP), and Union3.
All quoted constraints are given at the $68\%$ CL, except for the sterile-neutrino mass parameter, for which the $95\%$ CL upper limit is reported.
The final rows list the values of $\min(-\log \mathcal{L})$ for both the sterile-neutrino and corresponding baseline scenarios, together with $\Delta \chi^2_{\rm min}$ and $\Delta {\rm AIC}$.
}
\label{tab:sterile_cpl_lcdm}

\scriptsize
\setlength{\tabcolsep}{3.0pt}
\renewcommand{\arraystretch}{1.16}

\begin{adjustbox}{width=\textwidth,center}
\begin{tabular}{lcccccc}
\toprule
& \multicolumn{2}{c}{\textbf{CMB+BAO DESI DR2}}
& \multicolumn{2}{c}{\textbf{CMB+BAO DESI DR2+PP}}
& \multicolumn{2}{c}{\textbf{CMB+BAO DESI DR2+Union3}} \\
\cmidrule(lr){2-3}
\cmidrule(lr){4-5}
\cmidrule(lr){6-7}
\textbf{Parameter}
& \textbf{CPL + FTS}
& \textbf{$\Lambda$CDM + FTS}
& \textbf{CPL + FTS}
& \textbf{$\Lambda$CDM + FTS}
& \textbf{CPL + FTS}
& \textbf{$\Lambda$CDM + FTS} \\
\midrule

$10^{2}\omega_{b}$
& $2.309^{+0.014}_{-0.014}$
& $2.315^{+0.012}_{-0.012}$
& $2.312^{+0.013}_{-0.014}$
& $2.319^{+0.012}_{-0.013}$
& $2.311^{+0.013}_{-0.013}$
& $2.317^{+0.012}_{-0.013}$ \\

$\omega_{\rm cdm}$
& $0.13512^{+0.00113}_{-0.00108}$
& $0.13460^{+0.00083}_{-0.00065}$
& $0.13493^{+0.00096}_{-0.00094}$
& $0.13402^{+0.00102}_{-0.00089}$
& $0.13516^{+0.00103}_{-0.00101}$
& $0.13441^{+0.00084}_{-0.00072}$ \\

$100\theta_{s}$
& $1.03992^{+0.00028}_{-0.00028}$
& $1.03995^{+0.00028}_{-0.00027}$
& $1.03993^{+0.00028}_{-0.00028}$
& $1.04004^{+0.00029}_{-0.00027}$
& $1.03994^{+0.00028}_{-0.00028}$
& $1.03998^{+0.00028}_{-0.00028}$ \\

$\ln 10^{10}A_{s}$
& $3.094^{+0.014}_{-0.019}$
& $3.093^{+0.015}_{-0.015}$
& $3.091^{+0.014}_{-0.017}$
& $3.099^{+0.018}_{-0.018}$
& $3.0921^{+0.015}_{-0.016}$
& $3.095^{+0.015}_{-0.017}$ \\

$n_{s}$
& $0.9997^{+0.0036}_{-0.0035}$
& $1.0009^{+0.0034}_{-0.0035}$
& $1.0004^{+0.0036}_{-0.0035}$
& $1.0025^{+0.0036}_{-0.0034}$
& $0.9996^{+0.0033}_{-0.0036}$
& $1.0014^{+0.0035}_{-0.0034}$ \\

$\tau_{\rm reio}$
& $0.0624^{+0.0073}_{-0.0095}$
& $0.0622^{+0.0083}_{-0.0081}$
& $0.0611^{+0.0074}_{-0.0087}$
& $0.0668^{+0.0078}_{-0.0091}$
& $0.0616^{+0.0070}_{-0.0086}$
& $0.0636^{+0.0076}_{-0.0087}$ \\

\midrule

$m_{s}\,[\mathrm{eV}]$ $(95\%\,{\rm CL})$
& $<0.335$
& $<0.112$
& $<0.218$
& $<0.194$
& $<0.293$
& $<0.148$ \\

\midrule

$w_{0}$
& $-0.62^{+0.20}_{-0.16}$
& $-1~(\mathrm{fixed})$
& $-0.849^{+0.053}_{-0.047}$
& $-1~(\mathrm{fixed})$
& $-0.699^{+0.087}_{-0.091}$
& $-1~(\mathrm{fixed})$ \\

$w_{a}$
& $-1.11^{+0.51}_{-0.59}$
& $0~(\mathrm{fixed})$
& $-0.45^{+0.18}_{-0.20}$
& $0~(\mathrm{fixed})$
& $-0.89^{+0.31}_{-0.29}$
& $0~(\mathrm{fixed})$ \\

\midrule

$H_{0}\,[\mathrm{km\,s^{-1}\,Mpc^{-1}}]$
& $69.7^{+1.5}_{-2.0}$
& $73.73^{+0.24}_{-0.22}$
& $72.08^{+0.58}_{-0.60}$
& $73.45^{+0.25}_{-0.24}$
& $70.54^{+0.85}_{-0.87}$
& $73.63^{+0.26}_{-0.27}$ \\

$\Omega_{\rm m}$
& $0.331^{+0.019}_{-0.017}$
& $0.2923^{+0.0024}_{-0.0025}$
& $0.3075^{+0.0055}_{-0.0055}$
& $0.2947^{+0.0024}_{-0.0025}$
& $0.3227^{+0.0086}_{-0.0086}$
& $0.2932^{+0.0027}_{-0.0026}$ \\

$S_{8}$
& $0.844^{+0.015}_{-0.014}$
& $0.836^{+0.011}_{-0.0089}$
& $0.840^{+0.013}_{-0.012}$
& $0.827^{+0.013}_{-0.012}$
& $0.843^{+0.015}_{-0.012}$
& $0.834^{+0.011}_{-0.0097}$ \\

\midrule

$\min(-\log\mathcal{L})_{\rm FTS}$
& $1421.89$
& $1422.53$
& $2127.39$
& $2132.10$
& $1433.08$
& $1438.42$ \\

$\min(-\log\mathcal{L})_{\rm Standard}$
& $1401.68$
& $1407.30$
& $2108.01$
& $2114.19$
& $1412.91$
& $1422.05$ \\

$\Delta\chi^2_{\rm min}$
& $40.42$
& $30.46$
& $38.76$
& $35.82$
& $40.34$
& $32.74$ \\

$\Delta{\rm AIC}$
& $42.42$
& $32.46$
& $40.76$
& $37.82$
& $42.34$
& $34.74$ \\

\bottomrule
\end{tabular}
\end{adjustbox}
\end{table*}

\begin{figure*}[htpb!]
    \centering
    \includegraphics[width=0.48\textwidth]{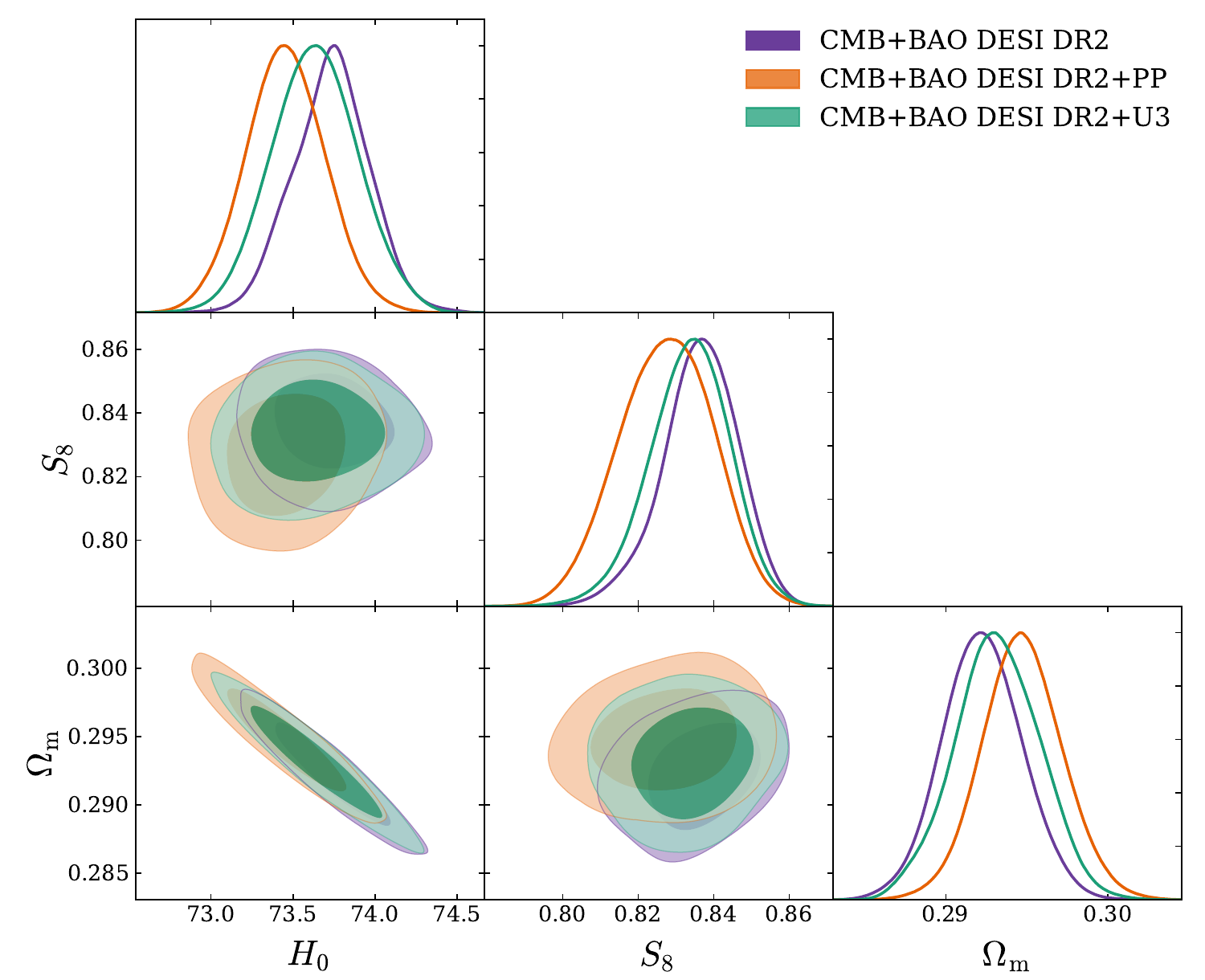} \,\, \,\,
    \includegraphics[width=0.48\textwidth]{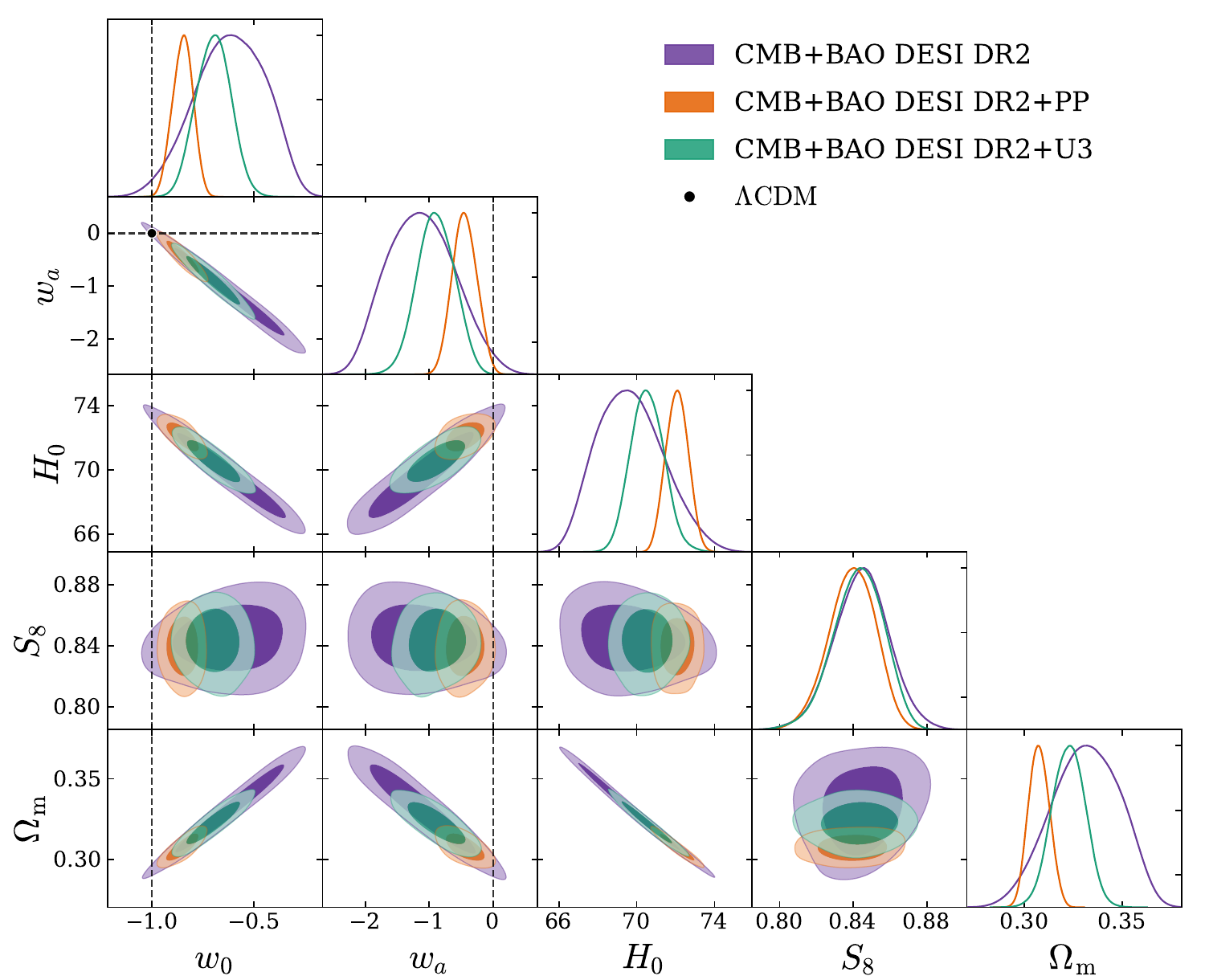}
\caption{
Marginalized one-dimensional posterior distributions and two-dimensional credible regions at the 68\% and 95\% CL for the $\Lambda$CDM (left panel) and CPL (right panel) cosmologies extended by the FTS sterile-neutrino scenario, for the data combinations indicated in the legend.
}
    \label{fig:FST_lcdm_cpl}
\end{figure*}

Table~\ref{tab:sterile_cpl_lcdm} summarizes the marginalized posterior constraints for the FTS scenario in both the standard $\Lambda$CDM framework and the CPL dark-energy extension, using combinations of Planck CMB data with DESI DR2 BAO and either the PantheonPlus or Union3 supernova datasets. Since this scenario assumes a fully populated sterile species with $\Delta N_{\rm eff}\simeq1$, it constitutes the most restrictive and observationally exposed realization considered in this work.

\textit{A first central result is that the FTS scenario is strongly disfavoured by current cosmological observations, independently of the assumed late-time cosmological background}. In the $\Lambda$CDM+FTS case, the minimum chi-square degradation relative to the corresponding active-neutrino reference model ranges from $\Delta\chi^2_{\rm min}\simeq30.5$ to $35.8$, depending on the dataset combination. The associated Akaike penalties are similarly large, with $\Delta {\rm AIC}\simeq32.5$--$37.8$. These values already indicate decisive statistical disfavouring of the fully thermalized sterile-neutrino hypothesis. In the CPL+FTS framework, the situation becomes even more severe, with $\Delta\chi^2_{\rm min}\simeq38.8$--$40.4$ and $\Delta{\rm AIC}\simeq40.8$--$42.4$. Therefore, allowing additional freedom in the dark-energy sector does not alleviate the cosmological tension generated by a fully thermalized sterile species. On the contrary, the larger parameter volume of CPL, combined with the poorer likelihood fit, further penalizes this realization.

These results demonstrate that current cosmological data strongly disfavour the assumption of complete sterile-neutrino thermalization. Importantly, this conclusion should not be interpreted as a generic rejection of all sterile-neutrino models, but rather as a specific constraint on the maximal-abundance thermal limit.

The inferred upper limits on the sterile-neutrino mass are correspondingly stringent. Within $\Lambda$CDM+FTS, the strongest bound is obtained from CMB+BAO DESI DR2,
\begin{equation}
m_s < 0.112~{\rm eV}\quad (95\%~{\rm CL}).
\end{equation}

In the CPL+FTS case, the corresponding bounds are moderately weakened,
\begin{equation}
m_s < 0.218\text{--}0.335~{\rm eV},
\end{equation}
depending on the dataset combination, reflecting the additional degeneracies introduced by the dark-energy parameters $w_0$ and $w_a$. Nevertheless, even in the most permissive case, eV-scale fully thermalized sterile neutrino masses remain decisively excluded.

A notable consequence of the fully thermalized sterile species is the systematic upward shift in the inferred Hubble constant. In the $\Lambda$CDM+FTS scenario, one finds
\begin{equation}
H_0 \simeq 73.4\text{--}73.7~{\rm km\,s^{-1}\,Mpc^{-1}},
\end{equation}
which is substantially larger than the canonical Planck $\Lambda$CDM value and significantly closer to local distance-ladder measurements~\cite{H0DN:2025lyy}. This behavior is expected because increasing $N_{\rm eff}$ raises the early expansion rate and reduces the sound horizon, thereby allowing a larger inferred value of $H_0$. In the CPL+FTS framework, the preferred Hubble constant spans a broader interval,
\begin{equation}
H_0 \simeq 69.7\text{--}72.1~{\rm km\,s^{-1}\,Mpc^{-1}},
\end{equation}
depending on the supernova compilation employed.

Although this shift does alleviate the Hubble tension, it is accompanied by a significant deterioration of the overall fit to the baseline datasets. This behavior contrasts with the dedicated analysis incorporating the local $H_0^{\rm DN}$ prior, where the $\Lambda$CDM+FTS model is able to reconcile the high local measurement of $H_0$ while remaining statistically comparable to the standard $\Lambda$CDM scenario when the full dataset, including SNIa observations, is considered.

The matter density parameter and clustering amplitude also exhibit systematic shifts. In the $\Lambda$CDM+FTS scenario, the preferred matter density remains near
\begin{equation}
\Omega_m \simeq 0.293,
\end{equation}
whereas in the CPL+FTS case it increases to
\begin{equation}
\Omega_m \simeq 0.31\text{--}0.33.
\end{equation}
This behavior is mainly driven by parameter degeneracies. Since CMB observations tightly constrain the physical matter density $\Omega_m h^2$, the lower values of $H_0$ preferred by the CPL+FTS model naturally shift the inferred value of $\Omega_m$ toward higher values compared to the $\Lambda$CDM+FTS scenario.

The clustering amplitude remains relatively high,
\begin{equation}
S_8 \simeq 0.83\text{--}0.84,
\end{equation}
indicating that the presence of a fully thermalized sterile neutrino tends to shift the inferred $S_8$ values toward even larger amplitudes compared to standard analyses without an additional sterile component.

The baseline cosmological parameters also display characteristic shifts associated with the additional relativistic degree of freedom. In particular, the scalar spectral index approaches near scale invariance,
\begin{equation}
n_s \simeq 1,
\end{equation}
while both $\omega_b$ and $\omega_{\rm cdm}$ are slightly increased relative to standard cosmology. These shifts reflect the well-known compensatory parameter adjustments required to preserve consistency with the CMB acoustic structure when $N_{\rm eff}$ is increased.

Comparing PantheonPlus and Union3, the inclusion of either supernova sample primarily sharpens late-time geometric constraints without qualitatively altering the overall conclusions. PantheonPlus generally yields slightly tighter constraints on some late-time cosmological parameters, while Union3 produces highly consistent results. This agreement confirms that the strong exclusion of the FTS scenario is robust against the specific choice of modern SNIa compilation.

Within the CPL parametrization, the analyses consistently favor a dynamical dark-energy sector with $w_{0}>-1$ and $w_{a}<0$. For the CMB+BAO DESI DR2 dataset, we obtain $w_{0}=-0.62^{+0.20}_{-0.16}$ and $w_{a}=-1.11^{+0.51}_{-0.59}$, while the inclusion of PantheonPlus tightens the constraints to $w_{0}=-0.849^{+0.053}_{-0.047}$ and $w_{a}=-0.45^{+0.18}_{-0.20}$. Using Union3 instead yields $w_{0}=-0.699^{+0.087}_{-0.091}$ and $w_{a}=-0.89^{+0.31}_{-0.29}$. In all cases, the negative values of $w_{a}$ indicate that the dark-energy equation of state evolves toward more negative values at earlier cosmic times, favoring a phantom-like behavior in the past, with $w(z\gg1)\simeq w_{0}+w_{a}<-1$. At the same time, the preference for $w_{0}>-1$ suggests a present-day equation of state less negative than a pure cosmological constant. This behavior points to a nontrivial redshift evolution of the dark-energy sector and may indicate a crossing of the phantom divide during the cosmic expansion history~\cite{Ozulker:2025ehg}. Moreover, the addition of supernova data considerably reduces the allowed parameter volume, especially in the PantheonPlus case, driving the solution closer to the $\Lambda$CDM limit while still preserving a mild preference for dynamical dark energy. We also note that this overall preference for $w_{0}>-1$ and $w_{a}<0$ is broadly consistent with the trends recently reported by the DESI collaboration in analyses combining BAO measurements with external cosmological datasets. In terms of the equivalent Gaussian-$\sigma$ deviation from the $\Lambda$CDM model, we find significances of $2.48\sigma$, $3.42\sigma$, and $4.09\sigma$ for the CMB+BAO DESI DR2, CMB+BAO DESI DR2+PP, and CMB+BAO DESI DR2+Union3 data combinations, respectively.

Despite the additional freedom introduced by the dark-energy parameters $w_{0}$ and $w_{a}$, which allow for significant departures from the standard $\Lambda$CDM scenario, the CPL+FTS model remains strongly disfavoured by current cosmological observations. This conclusion is supported by the large values of $\Delta \chi^{2}_{\rm min}$ and $\Delta {\rm AIC}$ reported in Table~\ref{tab:sterile_cpl_lcdm}. As will be discussed later, the same overall trend is further reinforced by the corresponding Bayesian model comparison through the Bayes factor.

Overall, the results of Table~\ref{tab:sterile_cpl_lcdm} indicate that a fully thermalized light sterile neutrino is strongly disfavoured by current cosmological observations. The combination of additional early radiation, modified acoustic features, and enhanced hot dark matter suppression produces a cosmological signature that is in strong tension with current high-precision datasets. While the model tends to increase the inferred value of $H_0$, this comes at the cost of a significantly poorer fit to the data.

Therefore, these cosmological data combinations do not support the fully thermalized sterile-neutrino hypothesis, either within the $\Lambda$CDM framework or in CPL cosmologies. This conclusion strongly motivates the exploration of less restrictive sterile-neutrino realizations, such as colder thermal relics or partially populated Dodelson--Widrow-like scenarios, in which the relation between physical mass, abundance, and cosmological observables is modified relative to the fully thermalized limit.

Figure~\ref{fig:FST_lcdm_cpl} presents the marginalized one-dimensional posterior distributions together with the corresponding two-dimensional credible regions at 68\% and 95\% CL for the fully thermalized sterile-neutrino (FTS) scenario within the $\Lambda$CDM (left panel) and CPL (right panel) cosmological frameworks, for the dataset combinations indicated in the legend. These contours visually illustrate the main conclusions discussed above, namely the strong cosmological disfavouring of the fully thermalized sterile-neutrino realization, the stringent upper bounds on the sterile mass, and the distinct parameter degeneracies that emerge when additional dark-energy freedom is introduced through the CPL extension.

\subsubsection{Impact of the Local $H_0^{\rm DN}$ Measurement on the $\Lambda$CDM+FTS Scenario}

We now assess how the FTS scenario responds to the explicit inclusion of a high local determination of the Hubble constant. For this purpose, we impose the Gaussian prior
\begin{equation}
H_0^{\rm DN}=73.5\pm0.81~{\rm km\,s^{-1}\,Mpc^{-1}},
\end{equation}
motivated by the recent Local Distance Network determination~\cite{H0DN:2025lyy}. This information is incorporated through importance reweighting of the existing $\Lambda$CDM+FTS and standard $\Lambda$CDM chains. Since no new dedicated MCMC analysis or likelihood minimization is performed, the resulting parameter constraints correspond to reweighted posterior estimates, while the associated goodness-of-fit and AIC quantities should be regarded as approximate indicators of the statistical impact of the local $H_0$ information.

\begin{table*}[t]
\centering
\caption{
Breakdown of the effective chi-square values for the $H_0^{\rm DN}$-reweighted analysis in the $\Lambda$CDM background. For each dataset combination, the $\Lambda$CDM+FTS scenario is compared against standard $\Lambda$CDM, with the same Gaussian prior $H_0^{\rm DN}=73.5\pm0.81~{\rm km\,s^{-1}\,Mpc^{-1}}$ applied to both models. The total effective chi-square is decomposed as $\chi^2_{\rm eff}=\chi^2_{\rm cosmo}+\chi^2_{H_0^{\rm DN}}$, where $\chi^2_{\rm cosmo}$ denotes the contribution from the cosmological datasets and $\chi^2_{H_0^{\rm DN}}$ the contribution from the local Gaussian prior. The quantity $H_0^{\rm bf}$ denotes the value of $H_0$ corresponding to the sample minimizing $\chi^2_{\rm eff}$ after importance reweighting. The last two columns report differences with respect to standard $\Lambda$CDM, following the convention $\Delta X=X(\Lambda{\rm CDM}+{\rm FTS})-X(\Lambda{\rm CDM})$.
}

\label{tab:h0dn_chi2_breakdown}
\footnotesize
\setlength{\tabcolsep}{4pt}
\renewcommand{\arraystretch}{1.16}

\resizebox{\textwidth}{!}{%
\begin{tabular}{l|l|c|c|c|c|c|c|c}
\hline\hline
\textbf{Dataset}
& \textbf{Scenario}
& \(H_0^{\rm bf}\)
& \(\boldsymbol{\chi^2_{\rm cosmo}}\)
& \(\boldsymbol{\chi^2_{H_0^{\rm DN}}}\)
& \(\boldsymbol{\chi^2_{\rm eff}}\)
& \(\boldsymbol{{\rm AIC}_{\rm eff}}\)
& \(\boldsymbol{\Delta\chi^2_{\rm eff}}\)
& \(\boldsymbol{\Delta{\rm AIC}_{\rm eff}}\)
\\
\hline\hline

CMB+BAO DESI DR2+\(H_0^{\rm DN}\)
& \(\Lambda\)CDM+FTS
& \(73.656\)
& \(2845.060\)
& \(0.037\)
& \(2845.097\)
& \(2901.097\)
& \(-5.100\)
& \(-3.100\)
\\
& Standard \(\Lambda\)CDM
& \(68.759\)
& \(2815.940\)
& \(34.257\)
& \(2850.197\)
& \(2904.197\)
& --
& --
\\
\hline

CMB+BAO DESI DR2+PP+\(H_0^{\rm DN}\)
& \(\Lambda\)CDM+FTS
& \(73.725\)
& \(4264.200\)
& \(0.077\)
& \(4264.277\)
& \(4322.277\)
& \(-1.338\)
& \(+0.662\)
\\
& Standard \(\Lambda\)CDM
& \(68.607\)
& \(4229.120\)
& \(36.495\)
& \(4265.615\)
& \(4321.615\)
& --
& --
\\
\hline

CMB+BAO DESI DR2+Union3+\(H_0^{\rm DN}\)
& \(\Lambda\)CDM+FTS
& \(73.692\)
& \(2876.840\)
& \(0.056\)
& \(2876.896\)
& \(2934.896\)
& \(-2.054\)
& \(-0.054\)
\\
& Standard \(\Lambda\)CDM
& \(68.869\)
& \(2846.260\)
& \(32.690\)
& \(2878.950\)
& \(2934.950\)
& --
& --
\\

\hline\hline
\end{tabular}}
\end{table*}

\begin{figure}[htpb!]
    \centering
    \includegraphics[width=0.5\textwidth]{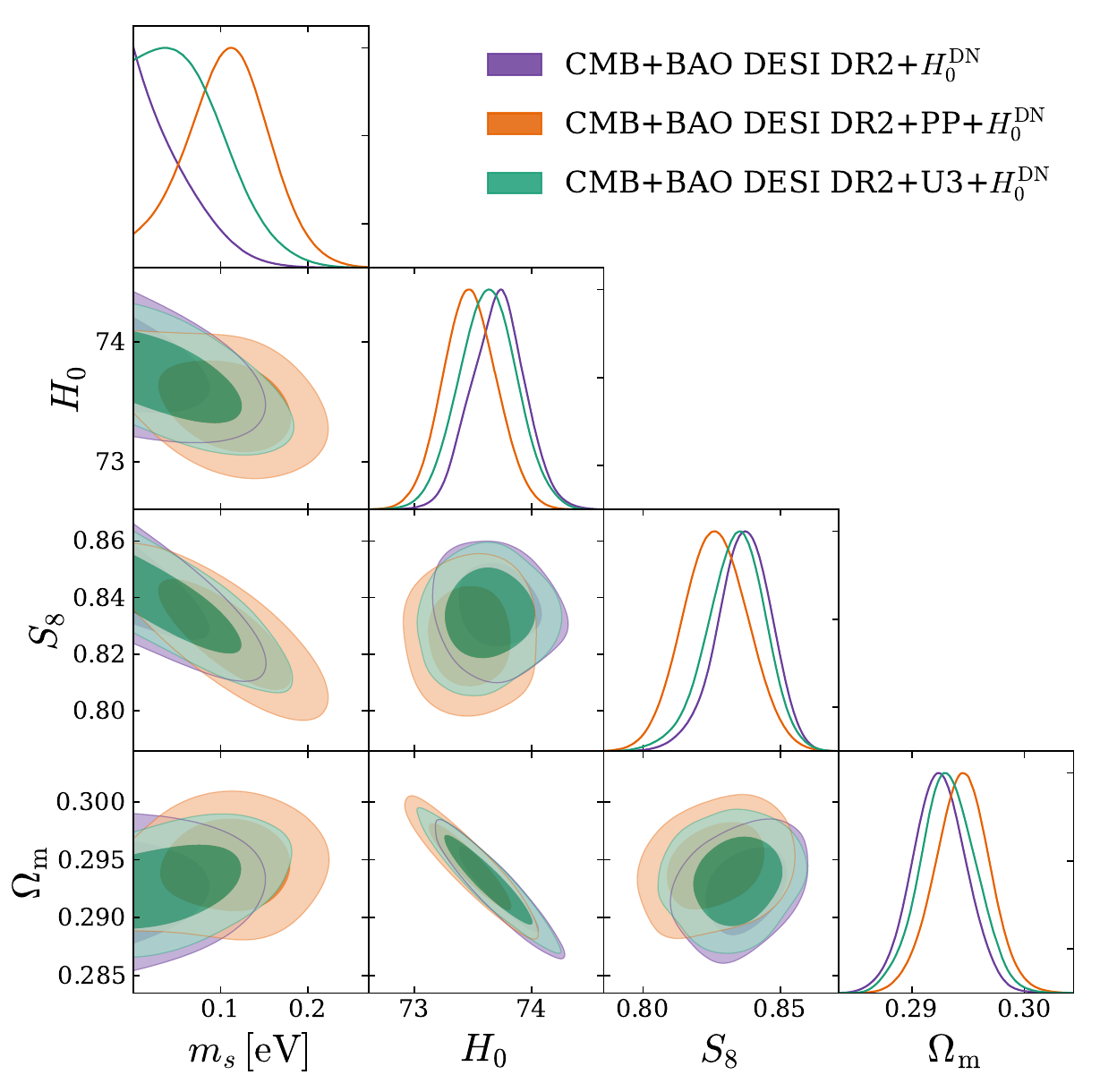}
\caption{
Marginalized one-dimensional posterior distributions and two-dimensional credible regions at the $68\%$ and $95\%$ CL for the $\Lambda$CDM cosmology extended by the fully thermalized sterile-neutrino scenario, including the Gaussian prior
$H_0^{\rm DN}=73.5\pm0.81~{\rm km\,s^{-1}\,Mpc^{-1}}$.
The contours correspond to the dataset combinations indicated in the legend.
}
\label{fig:lcdm_fts_h0dn}
\end{figure}

The corresponding reweighted constraints are shown in Fig.~\ref{fig:lcdm_fts_h0dn}, while the associated effective model-comparison estimates are reported in Table~\ref{tab:h0dn_chi2_breakdown}. The inclusion of the $H_0^{\rm DN}$ prior selects the high-$H_0$ region already present in the $\Lambda$CDM+FTS posterior, yielding values close to $H_0\simeq73.5~{\rm km\,s^{-1}\,Mpc^{-1}}$ for all dataset combinations. This behavior reflects the well-known degeneracy between the effective number of relativistic species and the Hubble constant, whereby the increased early-time expansion rate induced by the fully thermalized sterile neutrino reduces the sound horizon and shifts the preferred cosmological solution toward larger values of $H_0$.

The sterile-neutrino mass remains consistent with zero, with $95\%$ CL upper limits
\begin{equation}
m_s < (0.119,\ 0.205,\ 0.155)~{\rm eV},
\end{equation}
for CMB+BAO DESI DR2, CMB+BAO DESI DR2+PP, and CMB+BAO DESI DR2+Union3, respectively, in all cases including $H_0^{\rm DN}$. Therefore, while the inclusion of the local $H_0^{\rm DN}$ prior improves the compatibility of the FTS scenario with direct measurements of the expansion rate, it does not provide evidence for a nonzero sterile-neutrino mass.

All entries in Table~\ref{tab:h0dn_chi2_breakdown} are evaluated at the sample minimizing the effective chi-square after inclusion of the $H_0^{\rm DN}$ prior. The decomposition in Table~\ref{tab:h0dn_chi2_breakdown} clarifies the origin of the change in model comparison. Standard $\Lambda$CDM retains a better fit to the cosmological datasets alone, as indicated by the smaller values of $\chi^2_{\rm cosmo}$. However, it pays a large penalty from the local $H_0$ prior, with $\chi^2_{H_0^{\rm DN}}\simeq 33\text{--}36$. By contrast, $\Lambda$CDM+FTS shifts the preferred expansion rate to $H_0\simeq73.6\text{--}73.7~{\rm km\,s^{-1}\,Mpc^{-1}}$, reducing the local-$H_0$ contribution to $\chi^2_{H_0^{\rm DN}}<0.1$. Therefore, although the FTS extension worsens the fit to the cosmological datasets alone, this degradation is compensated by the removal of the local $H_0$ penalty.

As a result, the full $H_0^{\rm DN}$-reweighted comparison yields
\begin{equation}
\Delta\chi^2_{\rm eff}=(-5.10,\,-1.34,\,-2.05),
\end{equation}
for CMB+BAO DESI DR2, CMB+BAO DESI DR2+PP, and CMB+BAO DESI DR2+Union3, respectively. The corresponding AIC differences are
\begin{equation}
\Delta{\rm AIC}_{\rm eff}=(-3.10,\,+0.66,\,-0.05).
\end{equation}

Thus, once the local $H_0^{\rm DN}$ prior is included, $\Lambda$CDM+FTS provides a modest AIC improvement over standard $\Lambda$CDM for CMB+BAO DESI DR2 and becomes statistically comparable to standard $\Lambda$CDM once either SNIa compilation is included. In this sense, the fully thermalized sterile-neutrino scenario is able to accommodate the local $H_0^{\rm DN}$ determination within the $\Lambda$CDM background while remaining statistically competitive with the standard cosmological model. Nevertheless, the sterile-neutrino mass remains consistent with zero, so this improvement should not be interpreted as evidence for a nonzero fully thermalized sterile neutrino.

\subsection{Thermal relic with a different decoupling temperature}

Table~\ref{tab:thermal_relic_cpl_lcdm} presents the marginalized posterior constraints for the thermal relic sterile-neutrino scenario (DTS). Relative to the fully thermalized case, this framework introduces an additional degree of freedom through the sterile temperature ratio, thereby allowing the sterile abundance and relativistic energy density to be significantly suppressed while preserving a thermal momentum structure. As a result, this scenario provides a substantially broader and physically more flexible realization of sterile-neutrino cosmology.

\begin{table*}[!t]
\centering
\caption{
Same as Table~\ref{tab:sterile_cpl_lcdm}, but for the DTS sterile-neutrino scenario.
The sterile-neutrino mass parameter and $\Delta N_{\rm eff}^{\rm sterile}$ are reported as
95\% CL upper limits. The quantities $\Delta\chi^2_{\rm min}$ and $\Delta{\rm AIC}$
are computed with respect to the corresponding baseline CPL or $\Lambda$CDM cosmology
without the DTS sterile-neutrino extension.
}
\label{tab:thermal_relic_cpl_lcdm}

\scriptsize
\setlength{\tabcolsep}{3.0pt}
\renewcommand{\arraystretch}{1.18}

\begin{adjustbox}{width=\textwidth,center}
\begin{tabular}{lcccccc}
\toprule
& \multicolumn{2}{c}{\textbf{CMB+BAO DESI DR2}}
& \multicolumn{2}{c}{\textbf{CMB+BAO DESI DR2+PP}}
& \multicolumn{2}{c}{\textbf{CMB+BAO DESI DR2+Union3}} \\
\cmidrule(lr){2-3}
\cmidrule(lr){4-5}
\cmidrule(lr){6-7}
\textbf{Parameter}
& \textbf{CPL + DTS}
& \textbf{$\Lambda$CDM + DTS}
& \textbf{CPL + DTS}
& \textbf{$\Lambda$CDM + DTS}
& \textbf{CPL + DTS}
& \textbf{$\Lambda$CDM + DTS} \\
\midrule

$10^{2}\omega_{b}$
& $2.241^{+0.013}_{-0.013}$
& $2.260^{+0.012}_{-0.012}$
& $2.244^{+0.013}_{-0.013}$
& $2.257^{+0.012}_{-0.012}$
& $2.244^{+0.015}_{-0.014}$
& $2.260^{+0.013}_{-0.013}$ \\

$\omega_{\rm cdm}$
& $0.11950^{+0.00106}_{-0.00072}$
& $0.11712^{+0.00074}_{-0.00054}$
& $0.11889^{+0.00110}_{-0.00077}$
& $0.11729^{+0.00078}_{-0.00056}$
& $0.11925^{+0.00094}_{-0.00075}$
& $0.11724^{+0.00070}_{-0.00048}$ \\

$100\theta_{s}$
& $1.04189^{+0.00028}_{-0.00028}$
& $1.04212^{+0.00026}_{-0.00026}$
& $1.04196^{+0.00027}_{-0.00026}$
& $1.04212^{+0.00026}_{-0.00027}$
& $1.04190^{+0.00029}_{-0.00027}$
& $1.04212^{+0.00026}_{-0.00026}$ \\

$\ln(10^{10}A_{s})$
& $3.041^{+0.014}_{-0.015}$
& $3.0557^{+0.0130}_{-0.0159}$
& $3.045^{+0.014}_{-0.014}$
& $3.055^{+0.015}_{-0.015}$
& $3.044^{+0.013}_{-0.016}$
& $3.058^{+0.016}_{-0.016}$ \\

$n_{s}$
& $0.9664^{+0.0034}_{-0.0035}$
& $0.9726^{+0.0032}_{-0.0031}$
& $0.9673^{+0.0040}_{-0.0034}$
& $0.9719^{+0.0033}_{-0.0032}$
& $0.9679^{+0.0037}_{-0.0041}$
& $0.9731^{+0.0033}_{-0.0038}$ \\

$\tau_{\rm reio}$
& $0.0526^{+0.0070}_{-0.0073}$
& $0.0618^{+0.0063}_{-0.0077}$
& $0.0554^{+0.0065}_{-0.0074}$
& $0.0615^{+0.0073}_{-0.0073}$
& $0.0541^{+0.0063}_{-0.0079}$
& $0.0630^{+0.0077}_{-0.0077}$ \\

\midrule

$m_{s}\,[\mathrm{eV}]$ $(95\%\,{\rm CL})$
& $<4.608$
& $<4.716$
& $<4.591$
& $<4.529$
& $<4.411$
& $<4.567$ \\

$\Delta N_{\rm eff}^{\rm sterile}$ $(95\%\,{\rm CL})$
& $<0.0548$
& $<0.0351$
& $<0.0448$
& $<0.0265$
& $<0.0808$
& $<0.0869$ \\

$\textbf{$m_s^{\rm eff}$}[\mathrm{eV}]$ $(95\%\,{\rm CL})$
& $<0.2816$
& $<0.1468$
& $<0.2014$
& $<0.1575$
& $<0.2157$
& $<0.1624$ \\

\midrule

$w_{0}$
& $-0.57^{+0.12}_{-0.15}$
& $-1~(\mathrm{fixed})$
& $-0.815^{+0.057}_{-0.053}$
& $-1~(\mathrm{fixed})$
& $-0.677^{+0.070}_{-0.086}$
& $-1~(\mathrm{fixed})$ \\

$w_{a}$
& $-1.33^{+0.44}_{-0.35}$
& $0~(\mathrm{fixed})$
& $-0.67^{+0.20}_{-0.20}$
& $0~(\mathrm{fixed})$
& $-1.04^{+0.30}_{-0.22}$
& $0~(\mathrm{fixed})$ \\

\midrule

$H_{0}\,[\mathrm{km\,s^{-1}\,Mpc^{-1}}]$
& $65.1^{+1.3}_{-1.2}$
& $68.61^{+0.21}_{-0.21}$
& $67.45^{+0.54}_{-0.56}$
& $68.55^{+0.22}_{-0.23}$
& $66.08^{+0.75}_{-0.76}$
& $68.61^{+0.21}_{-0.26}$ \\

$\Omega_{\rm m}$
& $0.338^{+0.012}_{-0.015}$
& $0.2990^{+0.0024}_{-0.0024}$
& $0.3132^{+0.0053}_{-0.0053}$
& $0.2998^{+0.0027}_{-0.0025}$
& $0.3276^{+0.0075}_{-0.0086}$
& $0.2997^{+0.0026}_{-0.0024}$ \\

$S_{8}$
& $0.836^{+0.013}_{-0.011}$
& $0.8031^{+0.0092}_{-0.0075}$
& $0.8224^{+0.0122}_{-0.0082}$
& $0.8047^{+0.0093}_{-0.0081}$
& $0.828^{+0.013}_{-0.012}$
& $0.802^{+0.010}_{-0.010}$ \\

\midrule

$\min(-\log\mathcal{L})_{\rm DTS}$
& $1402.18$
& $1406.82$
& $2107.85$
& $2114.33$
& $1413.06$
& $1421.79$ \\

$\min(-\log\mathcal{L})_{\rm Standard}$
& $1401.68$
& $1406.98$
& $2108.01$
& $2114.19$
& $1412.91$
& $1422.05$ \\

$\Delta\chi^2_{\rm min}$
& $1.00$
& $-0.32$
& $-0.32$
& $0.28$
& $0.30$
& $-0.52$ \\

$\Delta{\rm AIC}$
& $5.00$
& $3.68$
& $3.68$
& $4.28$
& $4.30$
& $3.48$ \\

\bottomrule
\end{tabular}
\end{adjustbox}
\end{table*}

\textit{A key result emerging from Table~\ref{tab:thermal_relic_cpl_lcdm} is that, unlike the fully thermalized scenario, the thermal relic realization is not strongly disfavoured by current cosmological observations.}. Across all dataset combinations, the degradation in fit relative to the corresponding active-neutrino reference cosmology is dramatically reduced, and in some cases the DTS scenario even provides a marginal improvement in the best-fit likelihood. In the $\Lambda$CDM+DTS case, the best-fit likelihood differences remain extremely small,
\begin{equation}
\Delta \chi^2_{\rm min} \simeq -0.52 \text{ to } +0.28,
\end{equation}
while in the CPL+DTS framework one finds similarly mild shifts,
\begin{equation}
\Delta \chi^2_{\rm min} \simeq -0.32 \text{ to } +1.00.
\end{equation}

\begin{figure*}[htpb!] 
    \centering 
    \includegraphics[width=0.48\textwidth]{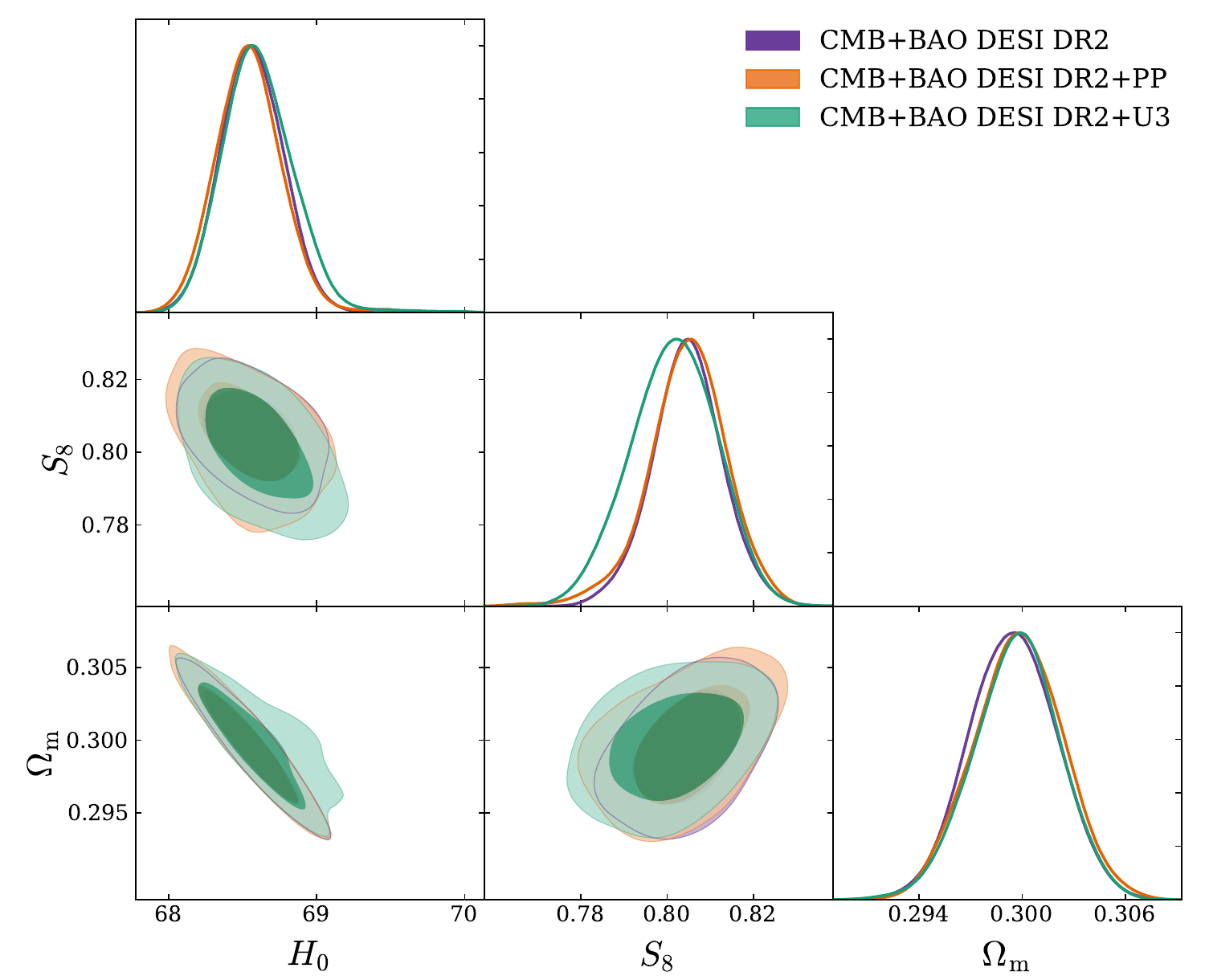} \,\, 
\,\,   
    \includegraphics[width=0.48\textwidth]{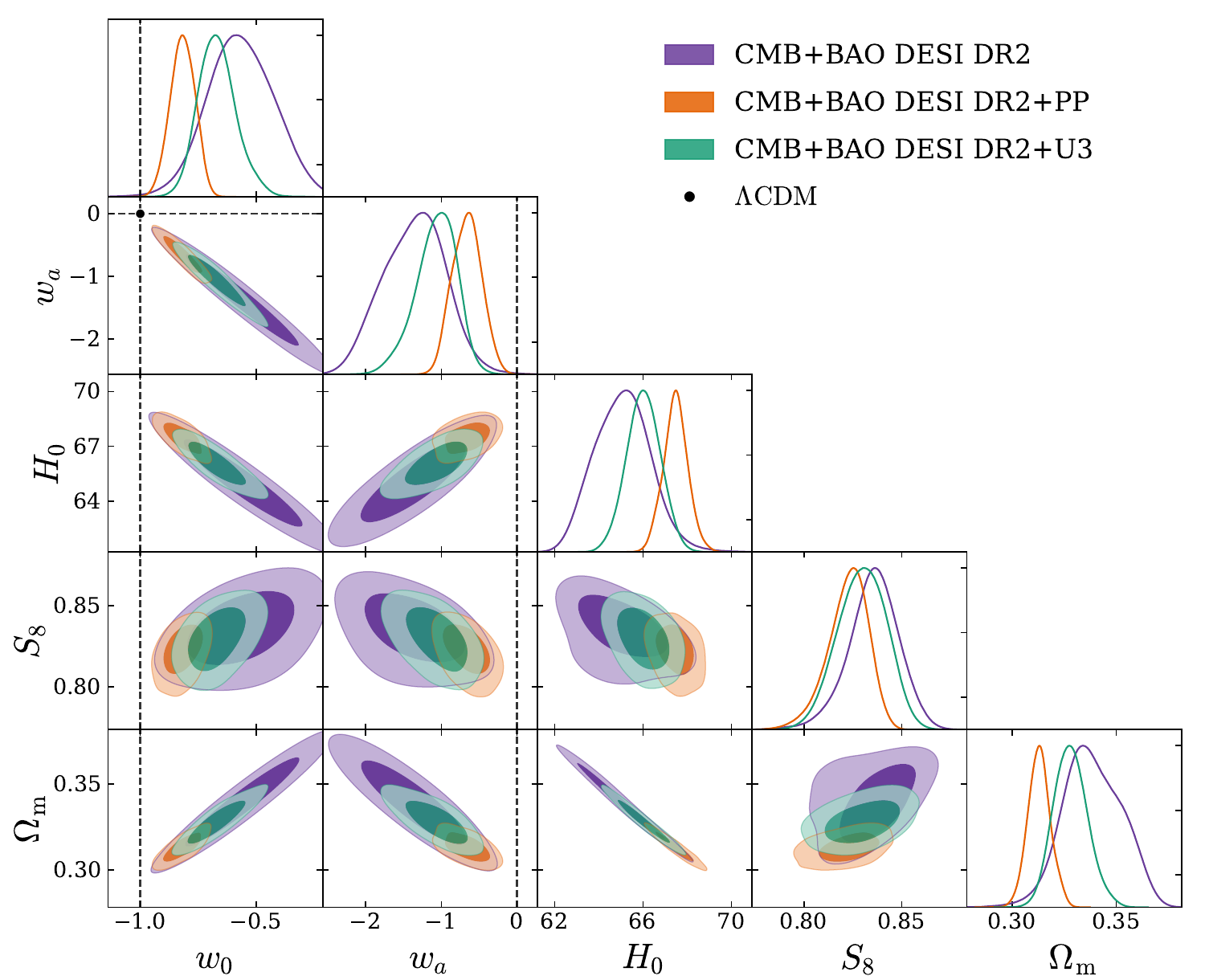} 
\caption{
Marginalized one-dimensional posterior distributions and two-dimensional credible regions at the 68\% and 95\% CL for the $\Lambda$CDM (left panel) and CPL (right panel) cosmologies extended by the DTS sterile-neutrino scenario, for the data combinations indicated in the legend.
}
\label{fig:DST_lcdm_cpl} 
\end{figure*}

These values indicate that current data neither strongly prefer nor decisively reject the DTS sterile-neutrino scenario at the level of goodness-of-fit alone. At the same time, the absence of a strong likelihood penalty should not be interpreted as evidence for a populated sterile sector. The posterior remains compatible with a negligible sterile abundance, with $\Delta N_{\rm eff}^{\rm sterile}$ consistent with zero and constrained to be small. Thus, the DTS scenario remains viable primarily because the data allow the sterile population to be strongly suppressed, effectively approaching the standard active-neutrino limit, rather than because they favor a nonzero sterile contribution.

This represents a major departure from the fully thermalized case. The cosmological exclusion is therefore not directed generically at sterile neutrinos themselves, but rather at the assumption of maximal thermal abundance. Once the sterile population is allowed to depart from the fully thermalized limit, its cosmological signatures can be sufficiently suppressed to evade the dramatic likelihood penalties observed in the FTS case.

The AIC, however, reveals a more nuanced picture. Although $\Delta\chi^2_{\rm min}$ is close to zero, the DTS scenario introduces an additional sterile-sector parameter and therefore incurs a complexity penalty. Consequently,
\begin{equation}
\Delta {\rm AIC} \simeq 3.5\text{--}5,
\end{equation}
depending on the cosmological background and dataset combination. This indicates mild to moderate statistical disfavouring, but not decisive exclusion. In practice, this means that current cosmological data do not require a thermal relic sterile neutrino, yet they also do not exclude it in any robust sense.

The sterile mass constraints are substantially weaker than in the FTS case. In $\Lambda$CDM+DTS, one finds
\begin{equation}
m_s < (4.716,\ 4.529,\ 4.567)~{\rm eV}
\qquad (95\%~{\rm CL}),
\end{equation}
while in CPL+DTS,
\begin{equation}
m_s < (4.608,\ 4.591,\ 4.411)~{\rm eV}
\qquad (95\%~{\rm CL}),
\end{equation}
for CMB+BAO DESI DR2, CMB+BAO DESI DR2+PP, and CMB+BAO DESI DR2+Union3, respectively. These limits allow significantly larger physical sterile masses than in the fully thermalized scenario.

However, the crucial point is that the cosmological impact is not governed directly by $m_s$, but rather by the effective parameters $\Delta N_{\rm eff}$ and \textbf{$m_s^{\rm eff}$}. The data strongly constrain the sterile abundance:
\begin{equation}
\Delta N_{\rm eff}^{\rm sterile} \lesssim 0.0265\text{--}0.0869,
\end{equation}
with the tightest limits generally arising in $\Lambda$CDM. Correspondingly, the effective sterile mass remains very small,
\begin{equation}
\textbf{$m_s^{\rm eff}$} \lesssim 0.1468\text{--}0.2816~{\rm eV}.
\end{equation}

These constraints imply that the cosmological viability of the DTS scenario is achieved primarily by suppressing the sterile abundance rather than by allowing large cosmological mass contributions. The corrected DTS constraints provide a direct illustration of the distinction between the physical sterile-neutrino mass and its cosmological impact. In the fully thermalized limit, one has $m_s^{\rm eff}=m_s$, so that a larger physical mass directly implies a larger hot-dark-matter contribution. In the DTS realization, instead, the relation in Eq.~\eqref{eq:meff_Ts} allows a comparatively large physical mass to coexist with a small effective mass whenever the sterile temperature, and hence its relic abundance, is sufficiently suppressed. Indeed, the present constraints permit $m_s$ to extend into the multi-eV regime within the adopted prior range, while simultaneously requiring $\Delta N_{\rm eff}^{\rm sterile}\lesssim0.09$ and $m_s^{\rm eff}\lesssim0.28~{\rm eV}$. Therefore, the viability of the DTS scenario does not arise from allowing a large cosmological hot-dark-matter density, but from diluting the sterile population so that its gravitational imprint remains limited. Since several of the physical-mass upper limits approach the adopted prior boundary, $m_s<5~{\rm eV}$, these results indicate that $m_s$ is only weakly constrained in the colder-relic scenario, rather than providing evidence for a preferred multi-eV sterile-neutrino mass.

From the perspective of late-time cosmological observables, the DTS scenario produces markedly different parameter shifts compared to FTS. In $\Lambda$CDM+DTS, the inferred Hubble constant becomes
\begin{equation}
H_0 \simeq 68.5\text{--}68.6~{\rm km\,s^{-1}\,Mpc^{-1}},
\end{equation}
while in CPL+DTS,
\begin{equation}
H_0 \simeq 65.1\text{--}67.5~{\rm km\,s^{-1}\,Mpc^{-1}}.
\end{equation}

Unlike the fully thermalized case, the DTS realization does not significantly raise $H_0$ toward local measurements. In fact, the preferred Hubble constant remains relatively close to standard cosmological values, particularly in $\Lambda$CDM. This reflects the strong upper bounds on $\Delta N_{\rm eff}$: because the sterile relic abundance is tightly suppressed, its effect on the sound horizon and early expansion history is correspondingly limited. Thus, the DTS scenario does not provide a compelling solution to the Hubble tension, but instead remains compatible with cosmological observations precisely because it behaves much more similarly to standard cosmology.

The matter density and clustering amplitude also show notable behavior. In the $\Lambda$CDM+DTS scenario, the preferred matter density is found to be $\Omega_m \simeq 0.300$, with $S_8 \simeq 0.80$. These values are moderately lower than those obtained in the FTS scenario. In the CPL+DTS framework, the inferred matter density increases to $\Omega_m \simeq 0.31\text{--}0.34$, while the clustering amplitude shifts to $S_8 \simeq 0.82\text{--}0.84$, reflecting the broader parameter degeneracies introduced by the dynamical dark-energy sector. This indicates that the DTS scenario can modestly suppress structure growth relative to the fully thermalized limit while avoiding the severe statistical penalties associated with FTS. Although the reduction in $S_8$ is not dramatic, it illustrates how partially suppressed sterile populations can alter late-time structure formation without introducing a significant degradation in the overall fit to the data.

At the level of primordial parameters, DTS cosmologies remain much closer to standard expectations than FTS. The scalar spectral index is found near $n_s \simeq 0.97$, rather than approaching scale invariance as in the fully thermalized case. Similarly, the baryon and cold dark matter densities remain close to canonical $\Lambda$CDM values. This reflects the relatively small allowed sterile abundance and demonstrates that DTS cosmologies largely preserve the standard acoustic structure of the CMB.

Comparing PantheonPlus and Union3 again reveals highly consistent results, with only minor quantitative differences. The addition of either supernova compilation primarily sharpens geometric constraints without significantly altering the sterile-sector conclusions. This confirms the robustness of the DTS parameter bounds across independent late-time probes.

An important aspect of the CPL+DTS results is the behavior of the dark-energy equation-of-state parameters, which reveal how the preferred late-time expansion history behaves in the presence of a suppressed sterile relic contribution. In all dataset combinations, the analyses consistently prefer $w_0>-1$ together with $w_a<0$, indicating a dynamical dark-energy component that evolves toward a more phantom-like regime at earlier times. Specifically, for CMB+BAO DESI DR2 we find $w_0=-0.57^{+0.12}_{-0.15}$ and $w_a=-1.33^{+0.44}_{-0.35}$, while the inclusion of PantheonPlus yields $w_0=-0.815^{+0.057}_{-0.053}$ and $w_a=-0.67^{+0.20}_{-0.20}$. Using Union3 instead gives $w_0=-0.677^{+0.070}_{-0.086}$ and $w_a=-1.04^{+0.30}_{-0.22}$. This behavior closely resembles the trends observed in the FTS case, although the DTS scenario generally permits slightly stronger dark-energy evolution due to the suppressed relativistic contribution of the sterile sector.

Taken together, these results lead to a fundamentally important conclusion: current cosmological data do not generically exclude thermal sterile relics with sufficiently suppressed abundances. The severe exclusions associated with the fully thermalized case arise specifically from the assumption of $\Delta N_{\rm eff}\simeq1$, rather than from the sterile-neutrino hypothesis itself. In this sense, the DTS scenario serves as a critical demonstration that sterile-neutrino viability is highly sensitive to thermal history. Current cosmological data strongly disfavour fully populated sterile sectors, but permit colder thermal relics whose contribution to the radiation density and hot dark matter sector remains subdominant.

Figure~\ref{fig:DST_lcdm_cpl} presents the marginalized posterior distributions of the principal baseline cosmological and sterile-sector parameters, together with the corresponding two-dimensional credible regions at 68\% and 95\% CL for the $\Lambda$CDM (left panel) and CPL (right panel) cosmologies extended by the DTS sterile-neutrino scenario. These contours highlight the dominant parameter correlations, degeneracy directions, and relative constraint strengths discussed above.

\subsection{Dodelson–Widrow-like sterile neutrino}

% ============================
% Table: DWLIKE for CPL and LCDM
% ============================

\begin{table*}[!t]
\centering
\caption{Same as Table~\ref{tab:sterile_cpl_lcdm}, but for the Dodelson--Widrow (DW)-like sterile-neutrino scenario.}
\label{tab:dwlike_cpl_lcdm}

\scriptsize
\setlength{\tabcolsep}{2.8pt}
\renewcommand{\arraystretch}{1.16}

\begin{adjustbox}{width=\textwidth,center}
\begin{tabular}{lcccccc}
\toprule
& \multicolumn{2}{c}{\textbf{CMB+BAO DESI DR2}}
& \multicolumn{2}{c}{\textbf{CMB+BAO DESI DR2+PP}}
& \multicolumn{2}{c}{\textbf{CMB+BAO DESI DR2+Union3}} \\
\cmidrule(lr){2-3}
\cmidrule(lr){4-5}
\cmidrule(lr){6-7}
\textbf{Parameter}
& \textbf{CPL + DW}
& \textbf{$\Lambda$CDM + DW}
& \textbf{CPL + DW}
& \textbf{$\Lambda$CDM + DW}
& \textbf{CPL + DW}
& \textbf{$\Lambda$CDM + DW} \\
\midrule

$10^{2}\omega_{b}$
& $2.244^{+0.015}_{-0.015}$
& $2.266^{+0.013}_{-0.015}$
& $2.251^{+0.015}_{-0.015}$
& $2.263^{+0.013}_{-0.014}$
& $2.248^{+0.014}_{-0.016}$
& $2.262^{+0.013}_{-0.013}$ \\

$\omega_{\rm cdm}$
& $0.1217^{+0.0013}_{-0.0018}$
& $0.1189^{+0.00044}_{-0.0018}$
& $0.1203^{+0.00079}_{-0.0015}$
& $0.1191^{+0.00033}_{-0.0020}$
& $0.1210^{+0.00102}_{-0.0015}$
& $0.1190^{+0.00054}_{-0.0018}$ \\

$100\theta_{s}$
& $1.04162^{+0.00033}_{-0.00035}$
& $1.04189^{+0.00038}_{-0.00028}$
& $1.04179^{+0.00032}_{-0.00029}$
& $1.04190^{+0.00039}_{-0.00027}$
& $1.04174^{+0.00034}_{-0.00028}$
& $1.04192^{+0.00039}_{-0.00028}$ \\

$\ln 10^{10}A_{s}$
& $3.042^{+0.014}_{-0.015}$
& $3.062^{+0.016}_{-0.018}$
& $3.049^{+0.014}_{-0.014}$
& $3.058^{+0.014}_{-0.014}$
& $3.046^{+0.016}_{-0.016}$
& $3.059^{+0.014}_{-0.014}$ \\

$n_{s}$
& $0.9692^{+0.0039}_{-0.0055}$
& $0.9754^{+0.0034}_{-0.0047}$
& $0.9709^{+0.0042}_{-0.0047}$
& $0.9749^{+0.0037}_{-0.0051}$
& $0.9698^{+0.0038}_{-0.0048}$
& $0.9748^{+0.0032}_{-0.0045}$ \\

$\tau_{\rm reio}$
& $0.0513^{+0.0072}_{-0.0068}$
& $0.0635^{+0.0075}_{-0.0088}$
& $0.0558^{+0.0067}_{-0.0067}$
& $0.0613^{+0.0071}_{-0.0071}$
& $0.0539^{+0.0072}_{-0.0080}$
& $0.0618^{+0.0067}_{-0.0066}$ \\

\midrule

$m_{s}\,[\mathrm{eV}]$ $(95\%\,{\rm CL})$
& $<0.550$
& $<0.548$
& $<0.411$
& $<0.678$
& $<0.593$
& $<0.626$ \\

$\Delta N_{\rm eff}^{\rm sterile}$ $(95\%\,{\rm CL})$
& $<0.2827$
& $<0.2847$
& $<0.2421$
& $<0.3119$
& $<0.2603$
& $<0.2864$ \\

$\textbf{$m_s^{\rm eff}$} \,[\mathrm{eV}]$ $(95\%\,{\rm CL})$
& $<0.1173$
& $<0.0593$
& $<0.0587$
& $<0.0403$
& $<0.1133$
& $<0.0337$ \\

\midrule

$w_{0}$
& $-0.700^{+0.16}_{-0.15}$
& $-1~(\mathrm{fixed})$
& $-0.833^{+0.059}_{-0.058}$
& $-1~(\mathrm{fixed})$
& $-0.651^{+0.082}_{-0.091}$
& $-1~(\mathrm{fixed})$ \\

$w_{a}$
& $-0.90^{+0.45}_{-0.46}$
& $0~(\mathrm{fixed})$
& $-0.60^{+0.21}_{-0.19}$
& $0~(\mathrm{fixed})$
& $-1.14^{+0.31}_{-0.27}$
& $0~(\mathrm{fixed})$ \\

\midrule

$H_{0}\,[\mathrm{km\,s^{-1}\,Mpc^{-1}}]$
& $66.6^{+1.6}_{-1.8}$
& $69.06^{+0.21}_{-0.58}$
& $67.89^{+0.65}_{-0.66}$
& $68.99^{+0.19}_{-0.70}$
& $66.31^{+0.83}_{-0.82}$
& $68.97^{+0.22}_{-0.58}$ \\

$\Omega_{\rm m}$
& $0.326^{+0.017}_{-0.016}$
& $0.2985^{+0.0027}_{-0.0025}$
& $0.3118^{+0.0056}_{-0.0057}$
& $0.2994^{+0.0029}_{-0.0026}$
& $0.3288^{+0.0086}_{-0.0083}$
& $0.2992^{+0.0026}_{-0.0026}$ \\

$S_{8}$
& $0.847^{+0.012}_{-0.012}$
& $0.8076^{+0.0078}_{-0.0094}$
& $0.8244^{+0.0092}_{-0.0092}$
& $0.8083^{+0.0079}_{-0.0083}$
& $0.833^{+0.010}_{-0.010}$
& $0.8088^{+0.0080}_{-0.0082}$ \\

\midrule

$\min(-\log\mathcal{L})_{\rm DW}$
& $1401.15$
& $1407.18$
& $2108.80$
& $2114.75$
& $1414.46$
& $1422.47$ \\

$\min(-\log\mathcal{L})_{\rm Standard}$
& $1401.68$
& $1407.30$
& $2108.01$
& $2114.19$
& $1412.91$
& $1422.05$ \\

$\Delta\chi^2_{\rm min}$
& $-1.06$
& $-0.24$
& $1.58$
& $1.12$
& $3.10$
& $0.84$ \\

$\Delta{\rm AIC}$
& $2.94$
& $3.76$
& $5.58$
& $5.12$
& $7.10$
& $4.84$ \\

\bottomrule
\end{tabular}
\end{adjustbox}
\end{table*}

\begin{figure*}[htpb!]
    \centering
    \includegraphics[width=0.48\textwidth]{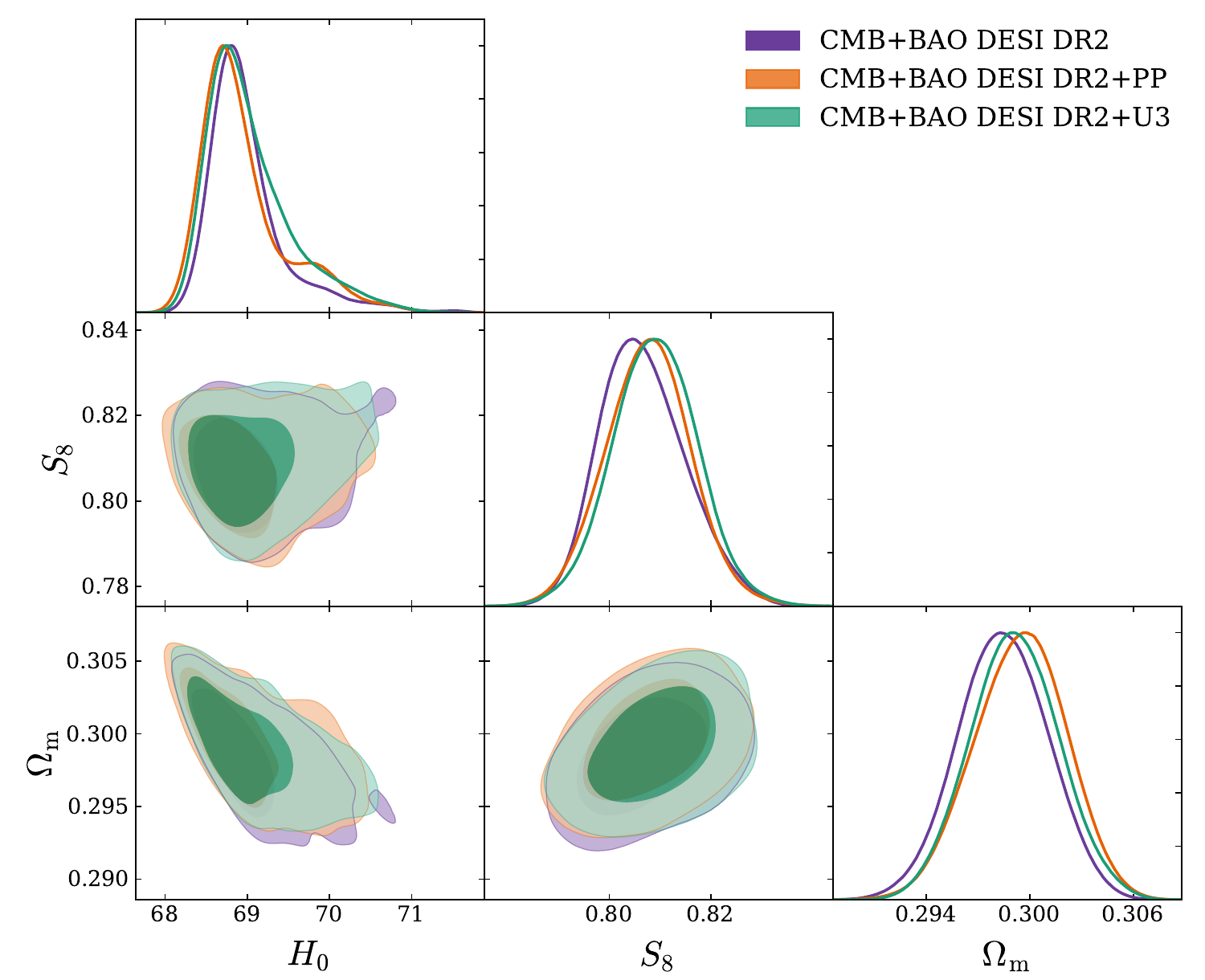} \,\, \,\,
    \includegraphics[width=0.48\textwidth]{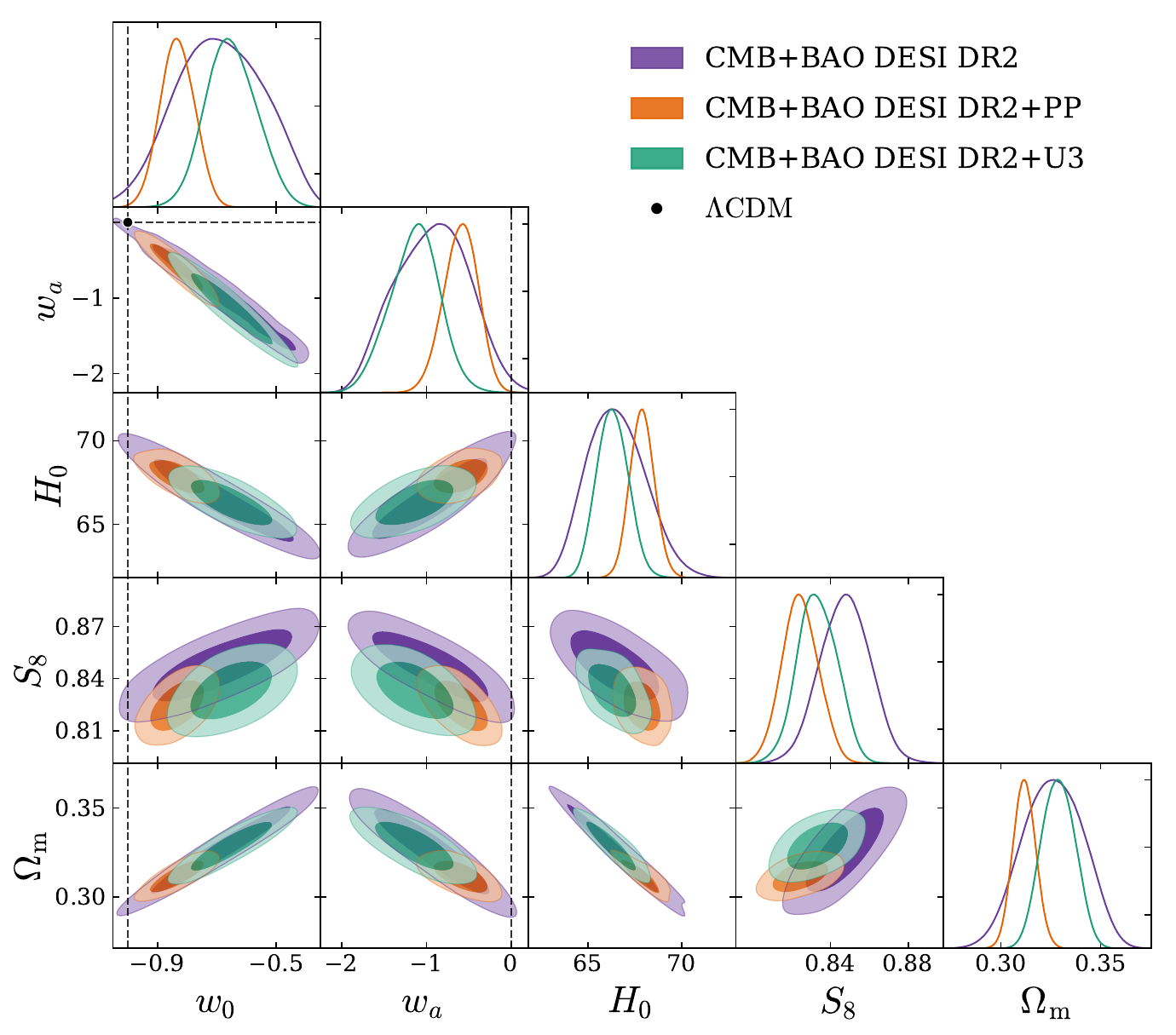}
\caption{
Same as Figure~\ref{fig:FST_lcdm_cpl}, but for the $\Lambda$CDM and CPL cosmologies extended by the DW sterile-neutrino scenario.
}
    \label{fig:DW_lcdm_cpl}
\end{figure*}

Table~\ref{tab:dwlike_cpl_lcdm} presents the marginalized posterior constraints for the Dodelson--Widrow-like (DW) sterile-neutrino scenario in both the $\Lambda$CDM and CPL cosmological frameworks. Relative to the thermal relic case, the DW scenario permits a different relation between the radiation density and effective sterile mass, leading to a distinct cosmological phenomenology.

\textit{A central conclusion from Table~\ref{tab:dwlike_cpl_lcdm} is that the DW-like scenario is substantially more viable than the fully thermalized limit, although it remains penalized once the additional sterile-sector freedom is taken into account.} In the $\Lambda$CDM+DW framework, the best-fit likelihood differences relative to the corresponding non-sterile reference cosmology remain small,
\begin{equation}
\Delta\chi^2_{\rm min} \simeq -0.24 \text{ to } +1.12,
\end{equation}
depending on the dataset combination. These values indicate that, in the standard $\Lambda$CDM background, the DW-like sterile-neutrino realization is not strongly disfavored at the level of goodness of fit.

The same conclusion broadly applies to the CPL+DW case after correcting the minimum-likelihood comparison. For the CMB+BAO DESI DR2 dataset alone, one finds
\begin{equation}
\Delta\chi^2_{\rm min} \simeq -1.06,
\end{equation}
rather than a large positive degradation. This behavior is both physically and statistically expected, since the DW-like scenario contains the corresponding baseline cosmology as the limiting case in which the sterile abundance becomes negligible. Once PantheonPlus or Union3 data are included, the goodness-of-fit shifts remain mild,
\begin{equation}
\Delta\chi^2_{\rm min} \simeq 1.6\text{--}3.1,
\end{equation}
showing that current data do not strongly reject the DW-like scenario at the best-fit level.

At the level of AIC, the additional sterile-sector freedom introduces the expected complexity penalty. In $\Lambda$CDM+DW, one finds
\begin{equation}
\Delta {\rm AIC} \simeq 3.8\text{--}5.1,
\end{equation}
while in CPL+DW the corrected values are
\begin{equation}
\Delta {\rm AIC} \simeq 2.9\text{--}7.1.
\end{equation}
Therefore, although the DW-like scenario is not strongly excluded in either $\Lambda$CDM or CPL cosmologies, it is not statistically preferred over the corresponding non-sterile reference model. As in the DTS case, its viability arises primarily because the sterile population can be driven toward negligible abundance, rather than because current data provide evidence for a nonzero sterile contribution.

The physical sterile-neutrino mass bounds are substantially tighter than those found for the thermal relic case,
\begin{equation}
m_s < 0.41\text{--}0.68~{\rm eV}
\qquad (95\%~{\rm CL}),
\end{equation}
depending on the cosmological background and dataset combination. However, as in all partially populated sterile scenarios, the cosmological impact is more directly determined by the effective abundance parameters.

A key distinction relative to the DTS scenario emerges in the constraints on $\Delta N_{\rm eff}^{\rm sterile}$. In the DW-like case, one finds substantially weaker limits,
\begin{equation}
\Delta N_{\rm eff}^{\rm sterile} \lesssim 0.24\text{--}0.31,
\end{equation}
which are considerably less restrictive than those obtained for the thermal relic case, where $\Delta N_{\rm eff}^{\rm sterile}$ was generally constrained below $\sim0.03$--$0.09$. This reflects the different mapping between abundance suppression and physical mass: because DW-like suppression occurs through the overall normalization of the distribution rather than through a reduced temperature, larger relativistic energy densities remain observationally allowed while preserving small effective mass contributions.

Part of this difference also arises from the prior structure of the two parameterizations. In the DTS scenario, the sampling is performed directly in the sterile temperature $T_s$, while $\Delta N_{\rm eff}^{\rm sterile}$ is a derived quantity related through Eq.~\eqref{eq:dneff_Ts}. As a consequence, a flat prior on $T_s$ induces a non-uniform prior on $\Delta N_{\rm eff}^{\rm sterile}$, assigning a larger fraction of the prior volume to low values of $\Delta N_{\rm eff}^{\rm sterile}$. This volume effect tends to pull the Bayesian credible intervals toward smaller values of $\Delta N_{\rm eff}^{\rm sterile}$ and therefore contributes to the tighter limits obtained in the DTS case. By contrast, in the DW scenario, $\Delta N_{\rm eff}^{\rm sterile}$ is sampled directly through the suppression factor $\chi$, leading to a different prior weighting of the allowed parameter space. Therefore, part of the difference between the DTS and DW constraints on $\Delta N_{\rm eff}^{\rm sterile}$ should be interpreted as a consequence of the different prior structures induced by the two physically motivated production mechanisms, in addition to any genuine differences in the cosmological response.

Despite this, the effective sterile mass remains tightly constrained:
\begin{equation}
\textbf{$m_s^{\rm eff}$} \lesssim 0.03\text{--}0.11~{\rm eV},
\end{equation}
which is generally tighter than the corresponding bounds obtained in the thermal relic scenario. Thus, while the DW scenario permits somewhat larger radiation-density contributions, the associated late-time hot-dark-matter contribution remains strongly constrained.

The inferred Hubble constant in the DW framework lies systematically above that of the colder thermal relic scenario, reflecting the larger allowed values of $\Delta N_{\rm eff}$. In the $\Lambda$CDM+DW scenario, the preferred value is $H_0 \simeq 69.0~{\rm km\,s^{-1}\,Mpc^{-1}}$, while in the CPL+DW framework the constraints shift to $H_0 \simeq 66.3\text{--}67.9~{\rm km\,s^{-1}\,Mpc^{-1}}$. These results indicate that the DW-like realization can modestly increase the inferred Hubble constant relative to standard cosmology, although much less efficiently than in the fully thermalized sterile-neutrino scenario. Consequently, the DW framework provides an intermediate phenomenology, allowing for a moderate enhancement in the early radiation density without incurring the strong statistical penalties associated with complete thermalization.

Similarly, the clustering amplitude is moderately reduced, with $\Lambda$CDM+DW preferring values around $S_8 \simeq 0.808$, somewhat lower than in the fully thermalized case and broadly comparable to the DTS scenario. At the level of the baseline cosmological parameters, the DW scenario remains relatively close to the standard cosmological model, although somewhat more shifted than the DTS case. The scalar spectral index typically falls within $n_s \simeq 0.97\text{--}0.975$, while both $\omega_b$ and $\omega_{\rm cdm}$ are mildly enhanced relative to baseline $\Lambda$CDM. These shifts are consistent with the moderate increase in relativistic energy density allowed in the DW framework.

The CPL dark-energy sector again exhibits a clear preference for dynamical dark energy, characterized by $w_0>-1$ and $w_a<0$ for all dataset combinations. For CMB+BAO DESI DR2, we obtain $w_0=-0.663^{+0.081}_{-0.083}$ and $w_a=-1.08^{+0.28}_{-0.26}$, while the inclusion of PantheonPlus shifts the constraints toward $w_0=-0.833^{+0.059}_{-0.058}$ and $w_a=-0.60^{+0.21}_{-0.19}$. Using Union3 instead yields $w_0=-0.651^{+0.082}_{-0.091}$ and $w_a=-1.14^{+0.31}_{-0.27}$. As in the FTS and DTS scenarios, the negative values of $w_a$ imply that the dark-energy equation of state evolves toward more negative values at earlier times, with $w(z\gg1)\simeq w_0+w_a<-1$, while the present-day value remains less negative than a pure cosmological constant. Overall, the DW scenario continues to exhibit a preference for a nontrivial late-time evolution of the dark-energy sector, with trends qualitatively consistent with those reported by the DESI collaboration in recent dynamical dark-energy analyses.

Taken together, the DW-like results reinforce a central message of this work: cosmological constraints on sterile neutrinos depend critically on the assumed production mechanism. While fully thermalized sterile species are strongly disfavored, partially populated DW-like realizations remain broadly viable provided that the effective sterile abundance is sufficiently suppressed.

Relative to the thermal relic scenario, the DW-like realization shows somewhat less restrictive marginalized bounds on $\Delta N_{\rm eff}^{\rm sterile}$, together with slightly larger shifts in parameters such as $H_0$, while leading to comparable limits on the effective sterile mass. This difference should not be overinterpreted as a purely physical preference for a larger sterile abundance in the DW-like case, since the DTS and DW-like scenarios are sampled using different production-level parameters and therefore correspond to different prior measures in the derived $\{\Delta N_{\rm eff},m_s^{\rm eff}\}$ plane. With this caveat, the DW scenario provides an important intermediate benchmark between the strongly disfavored fully thermalized limit and colder thermal relic realizations with suppressed abundance.

Figure~\ref{fig:DW_lcdm_cpl} displays the marginalized posterior distributions for the principal baseline cosmological and sterile-sector parameters, together with the corresponding two-dimensional credible regions at the 68\% and 95\% CL, for the $\Lambda$CDM (left panel) and CPL (right panel) cosmological frameworks extended by the DW sterile-neutrino scenario.

\subsection{Global Comparison Among All Cases}

% ============================
% FINAL Table: Bayesian evidence
% CPL in black, LCDM in blue
% ============================
\begin{table}[htpb!]
\centering
\caption{
Differences in Bayes factors, $\Delta \ln B$, for the sterile-neutrino scenarios considered. Black entries correspond to the CPL cosmology, blue entries correspond to the $\Lambda$CDM cosmology, and red entries show the importance-reweighted Bayes factors for $\Lambda$CDM+FTS after including the local $H_0^{\rm DN}=73.5\pm0.81~{\rm km\,s^{-1}\,Mpc^{-1}}$ prior, relative to standard $\Lambda$CDM with the same prior. In each case, the sterile-neutrino extension is compared with the corresponding baseline cosmology without the sterile component.
}
\label{tab:bayes_factor_sterile}

\small
\setlength{\tabcolsep}{3.2pt}
\renewcommand{\arraystretch}{1.20}

\begin{tabular*}{\columnwidth}{@{\extracolsep{\fill}}c|c|c|c|c@{}}
\hline\hline
\textbf{Dataset} &
\textbf{Case} &
\textbf{CPL} &
\textcolor{blue}{\textbf{\(\Lambda\)CDM}} &
\textcolor{red}{\textbf{\(+H_0^{\rm DN}\)}} \\
\hline\hline

& FTS
& $-22.01$
& \textcolor{blue}{$-18.38$}
& \textcolor{red}{$-1.18$} \\
\begin{tabular}[c]{@{}c@{}}CMB+BAO\\DESI DR2\end{tabular}
& DW
& $ -3.81$
& \textcolor{blue}{$-3.23$}
& \textcolor{red}{--} \\
& DTS
& $-2.75$
& \textcolor{blue}{$-0.93$}
& \textcolor{red}{--} \\
\hline

& FTS
& $-19.64$
& \textcolor{blue}{$-19.31$}
& \textcolor{red}{$-0.23$} \\
\begin{tabular}[c]{@{}c@{}}CMB+BAO\\DESI DR2+PP\end{tabular}
& DW
& $-2.69$
& \textcolor{blue}{$-3.37$}
& \textcolor{red}{--} \\
& DTS
& $0.82$
& \textcolor{blue}{$-0.21$}
& \textcolor{red}{--} \\
\hline

& FTS
& $-20.10$
& \textcolor{blue}{$-17.96$}
& \textcolor{red}{$+0.08$} \\
\begin{tabular}[c]{@{}c@{}}CMB+BAO\\DESI DR2+Union3\end{tabular}
& DW
& $-2.90$
& \textcolor{blue}{$-1.81$}
& \textcolor{red}{--} \\
& DTS
& $0.95$
& \textcolor{blue}{$0.56$}
& \textcolor{red}{--} \\
\hline\hline
\end{tabular*}
\end{table}

\begin{figure*}[htpb!]
    \centering
    \includegraphics[width=0.6\textwidth]{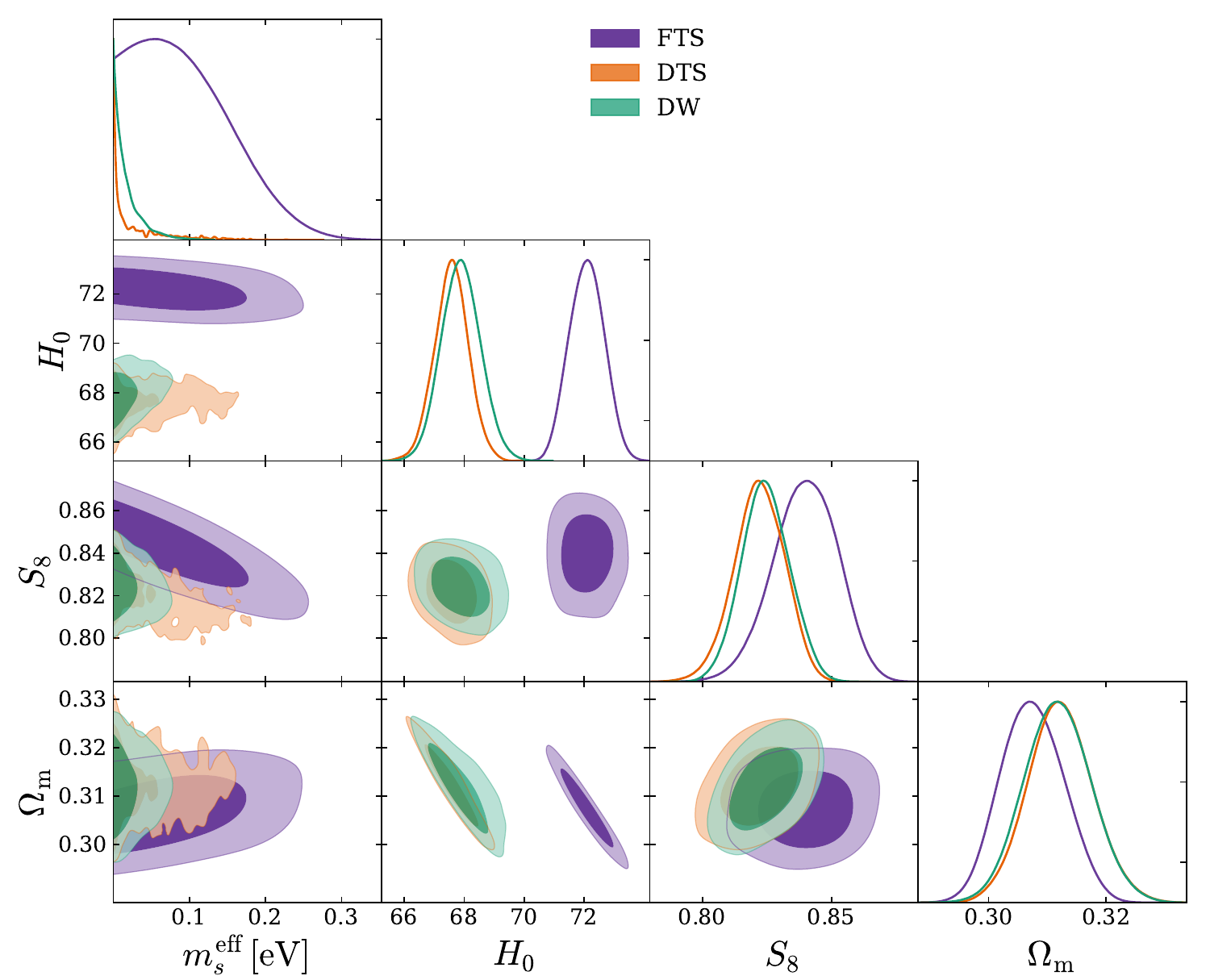}
   \caption{
Marginalized one-dimensional posterior distributions and two-dimensional credible regions at 68\% and 95\% CL for the $\Lambda$CDM cosmology extended by the FTS, DTS, and DW sterile-neutrino scenarios, obtained from the joint analysis of the CMB+BAO DESI DR2+PP dataset.
}
    \label{fig:global_lcdm}
\end{figure*}

\begin{figure*}[htpb!]
    \centering
    \includegraphics[width=0.6\textwidth]{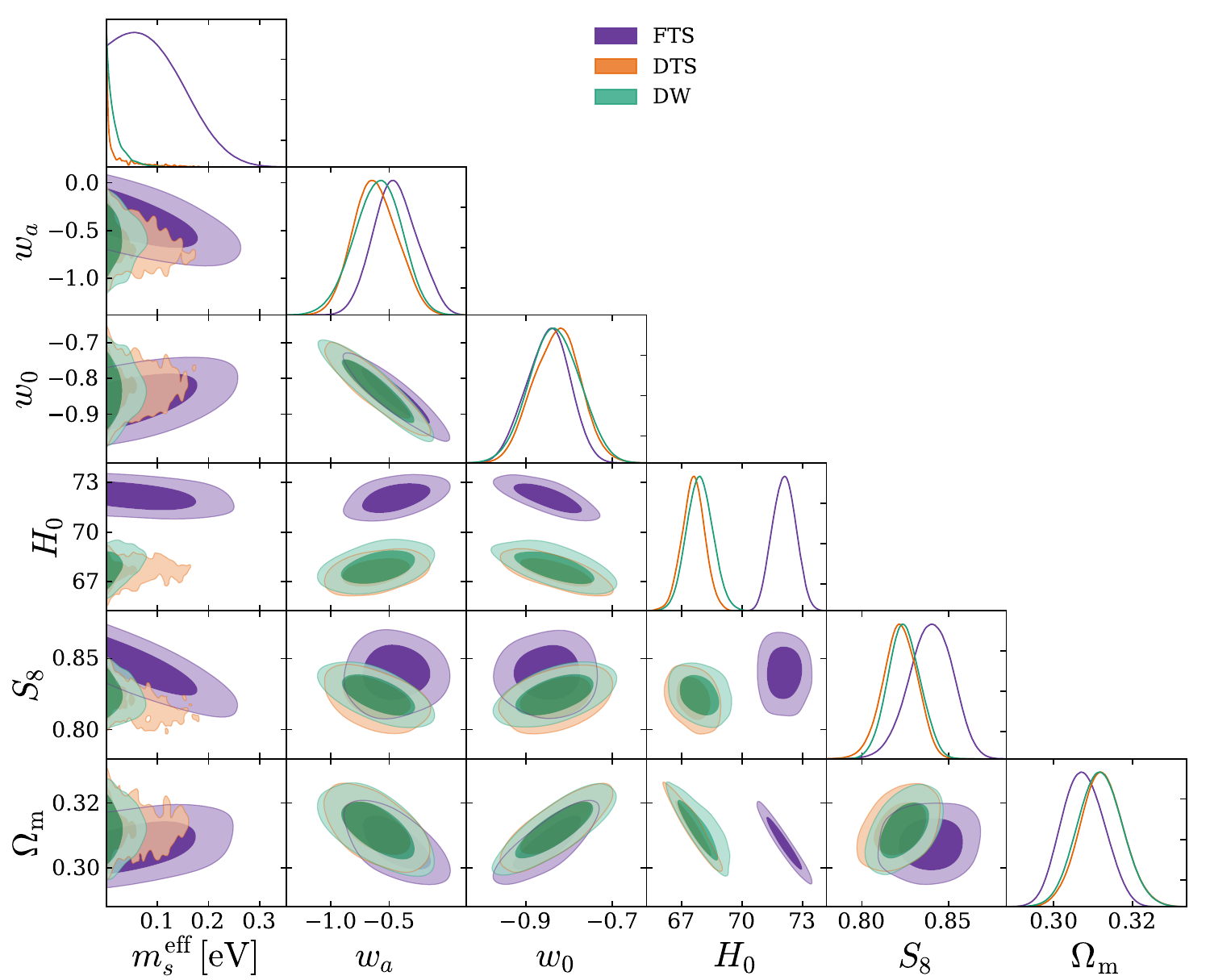}
\caption{
Same as Figure~\ref{fig:global_lcdm}, but for the CPL cosmology extended by the FTS, DTS, and DW sterile-neutrino scenarios.
}
    \label{fig:global_cpl}
\end{figure*}

Table~\ref{tab:bayes_factor_sterile} summarizes the Bayesian evidence differences, $\Delta \ln B$, for the three sterile-neutrino realizations considered in this work, within both the CPL and $\Lambda$CDM cosmological backgrounds. The Bayesian evidence results provide a clear statistical demonstration of the central thesis of this work: current cosmological data do not generically exclude sterile neutrinos, but rather strongly disfavour the specific assumption of full thermalization.

The fully thermalized sterile-neutrino scenario is decisively disfavored across all cosmological backgrounds and dataset combinations. For the FTS case, one finds
\begin{equation}
\Delta \ln B \sim -18 \text{ to } -22,
\end{equation}
for both $\Lambda$CDM and CPL. Such strongly negative Bayes factors correspond to overwhelming Bayesian evidence against the fully thermalized scenario in the baseline analyses without a local $H_0^{\rm DN}$ prior. Importantly, this conclusion is remarkably robust to the inclusion of supernova datasets and to the choice of late-time cosmological background. Whether one considers CMB+BAO DESI DR2 alone or adds PantheonPlus or Union3, the evidence remains consistently and strongly negative. This demonstrates that the cosmological exclusion of the FTS scenario is not driven by a particular dataset combination or dark-energy parameterization, but is instead a generic consequence of the large radiation density and hot-dark-matter contribution implied by full thermal equilibrium.

The last column of Table~\ref{tab:bayes_factor_sterile} shows that this Bayesian conclusion changes substantially once the local $H_0^{\rm DN}$ prior is included in the $\Lambda$CDM+FTS comparison. After importance reweighting, the Bayes factors become $\Delta\ln B_{H_0^{\rm DN}}=(-1.18,-0.23,+0.08)$ for CMB+BAO DESI DR2, CMB+BAO DESI DR2+PP, and CMB+BAO DESI DR2+Union3, respectively. Thus, the overwhelming Bayesian evidence against FTS found in the baseline analysis is not preserved when high local-$H_0$ information is imposed. Instead, the evidence becomes weakly negative for CMB+BAO DESI DR2 and statistically inconclusive once either SNIa compilation is included.
This behavior is consistent with the $H_0^{\rm DN}$-reweighted $\chi^2_{\rm eff}$ and AIC analysis above: the local $H_0$ prior makes $\Lambda$CDM+FTS statistically competitive with standard $\Lambda$CDM, while simultaneously shifting the inferred expansion rate into excellent agreement with the local determination. Therefore, if the $H_0^{\rm DN}$ measurement accurately reflects the true value of the Hubble constant, the fully thermalized sterile-neutrino scenario provides a viable resolution of the Hubble tension within the $\Lambda$CDM framework. Nevertheless, the sterile-neutrino mass remains compatible with zero, indicating that the preference arises primarily from the additional relativistic energy density associated with the fully thermalized sterile sector rather than from evidence for a massive sterile state.

By contrast, the DW-like scenario exhibits substantially weaker Bayesian penalties than the fully thermalized case. In $\Lambda$CDM, the DW case yields
\begin{equation}
\Delta \ln B \simeq -3.23,\,-3.37,\,-1.81,
\end{equation}
for CMB+BAO DESI DR2, CMB+BAO DESI DR2+PP, and CMB+BAO DESI DR2+Union3, respectively. These values indicate weak-to-moderate Bayesian evidence against the model, with the statistical penalty significantly reduced relative to FTS.

In CPL, the corresponding DW values are
\begin{equation}
\Delta \ln B \simeq -3.81,\,-2.69,\,-2.90,
\end{equation}
for the same dataset combinations. Therefore, the CMB+BAO DESI DR2-only combination does not produce a decisive Bayesian rejection of DW-like sterile neutrinos, but rather a moderate disfavouring. The inclusion of supernova data mildly improves the Bayesian viability of the CPL+DW scenario, although the evidence remains negative. Overall, DW-like sterile neutrinos are statistically disfavored, but far less severely than the fully thermalized limit.

The DTS scenario provides the least penalized case among the sterile-neutrino realizations considered here. In $\Lambda$CDM,
\begin{equation}
\Delta \ln B \simeq -0.93,\,-0.21,\,+0.56,
\end{equation}
while in CPL,
\begin{equation}
\Delta \ln B \simeq -2.75,\,+0.82,\,+0.95.
\end{equation}

These values imply the following conclusions:

\begin{enumerate}
\item For CMB+BAO DESI DR2 alone, DTS remains disfavored, although much less severely than FTS and generally less than DW.
\item Once supernova data are included, the Bayesian evidence becomes statistically inconclusive or mildly favorable, depending on the cosmological background and SNe compilation.
\item For the PantheonPlus combination, DTS is mildly favored in CPL but remains nearly neutral, though slightly negative, in $\Lambda$CDM. For the Union3 combination, DTS becomes mildly favored in both CPL and $\Lambda$CDM.
\end{enumerate}

These results indicate that, when the sterile abundance is sufficiently suppressed through a reduced thermal temperature, current cosmological observations permit sterile neutrinos and can even exhibit a mild Bayesian preference for them in some dataset combinations. However, this preference remains below the threshold conventionally regarded as significant. From a broader statistical perspective, these Bayesian evidence patterns establish a clear hierarchy:
\begin{equation}
{\rm FTS} \ll {\rm DW} < {\rm DTS}.
\end{equation}

This hierarchy directly reflects the underlying physical assumptions:

\begin{itemize}
\item FTS maximizes both $\Delta N_{\rm eff}$ and the late-time sterile abundance;
\item DW partially suppresses the sterile abundance through a reduced normalization of the distribution function;
\item DTS suppresses the cosmological signatures of the sterile sector through a colder thermal distribution and the corresponding reduction in relic abundance.
\end{itemize}

Therefore, the Bayesian evidence strongly confirms that cosmological constraints are primarily excluding highly populated sterile sectors rather than sterile neutrinos as a general physical category.

An additional important observation is the relative behavior of CPL versus $\Lambda$CDM. The impact of allowing a dynamical dark-energy sector is not uniform across all sterile-neutrino scenarios and datasets. For DTS, CPL improves the Bayesian viability once SNe data are included, leading to mildly positive values of $\Delta\ln B$ for both PantheonPlus and Union3. However, for CMB+BAO DESI DR2 alone, the $\Lambda$CDM+DTS case is less penalized than CPL+DTS. Similarly, for DW the CPL extension does not systematically improve the evidence relative to $\Lambda$CDM. Thus, dark-energy freedom can help absorb sterile-sector effects in some late-time dataset combinations, but it is not by itself sufficient to guarantee improved Bayesian support.

Taken together, Table~\ref{tab:bayes_factor_sterile} provides perhaps the most comprehensive statistical summary of this work. The results demonstrate that:

\begin{itemize}
\item Fully thermalized sterile neutrinos are decisively disfavored by current cosmological observations in the baseline analyses. Nevertheless, when a local $H_0$ prior is included, $\Lambda$CDM+FTS becomes statistically competitive with standard $\Lambda$CDM, especially once SNIa data are included;
\item DW-like sterile neutrinos are moderately disfavored, but remain far more viable than the fully thermalized limit;
\item Thermal relic sterile neutrinos with suppressed temperature are broadly allowed and may even be mildly favored for some dataset combinations, especially when SNe data are included.
\end{itemize}

These results have broad implications for the interpretation of cosmological neutrino constraints. Statements claiming that cosmology excludes sterile neutrinos are only accurate when applied to the fully thermalized benchmark. More general partially populated sterile-neutrino realizations remain consistent with current cosmological observations and, in some cases, exhibit a mild Bayesian preference relative to the corresponding baseline cosmology.

Figures~\ref{fig:global_lcdm} and \ref{fig:global_cpl} show the marginalized one-dimensional posterior distributions and two-dimensional credible regions at 68\% and 95\% CL for the $\Lambda$CDM and CPL cosmologies, respectively. The figures provide a direct comparison among all sterile-neutrino scenarios investigated in this work using the combined CMB+DESI DR2+PantheonPlus dataset. Overall, the main parameter degeneracies and cosmological trends discussed above can be clearly identified in the joint posterior distributions. Similar qualitative conclusions are obtained for the remaining dataset combinations analyzed throughout this work.

\section{Conclusion}
\label{final}

In this work, we have presented a detailed cosmological investigation of light sterile-neutrino scenarios aimed at clarifying a central question in contemporary cosmology: whether present cosmological observations exclude sterile neutrinos in general, or instead primarily disfavour the highly restrictive assumption of full thermalization.

To address this question, we considered three physically motivated sterile-neutrino realizations spanning a broad range of thermal histories and phase-space structures: the fully thermalized sterile neutrino (FTS), the thermal relic scenario with a reduced sterile temperature (DTS), and the Dodelson--Widrow-like (DW) realization with suppressed phase-space normalization. These scenarios were analyzed consistently within both the standard $\Lambda$CDM cosmology and the CPL dynamical dark-energy extension, allowing us to disentangle the effects of sterile-sector physics from assumptions regarding the late-time cosmic expansion history.

The results reveal a clear hierarchy among sterile-neutrino realizations, although the interpretation of this hierarchy must be tied to the production mechanism and prior prescription adopted for each scenario.

First, for the baseline cosmological combinations without a local $H_0$ prior, the fully thermalized sterile-neutrino scenario is strongly disfavored across all dataset combinations and cosmological backgrounds. This disfavouring is driven by the unavoidable combination of maximal early-time radiation density, $\Delta N_{\rm eff}\simeq1$, and substantial late-time hot-dark-matter abundance, which together produce cosmological signatures incompatible with current precision observations. However, the supplementary $H_0^{\rm DN}$-reweighted analysis shows that this conclusion is substantially weakened once high local values of $H_0$ are explicitly imposed. Within $\Lambda$CDM, the FTS scenario can accommodate the local $H_0^{\rm DN}$ determination and becomes statistically comparable to standard $\Lambda$CDM once SNIa data are included. Therefore, if the local $H_0^{\rm DN}$ measurement accurately reflects the true value of the Hubble constant, the fully thermalized sterile-neutrino scenario provides a viable resolution of the Hubble tension within the $\Lambda$CDM framework. This improvement, however, does not correspond to evidence for a nonzero sterile-neutrino mass, which remains compatible with zero.

Second, the DW-like scenario substantially relaxes the tensions found in the fully thermalized case. By allowing a suppressed sterile abundance through incomplete population, this framework remains broadly compatible with current cosmological data, particularly within $\Lambda$CDM. Although still penalized by information criteria because of the additional sterile-sector parameters, the DW realization is far less statistically disfavored than the fully thermalized case, demonstrating that cosmological constraints are highly sensitive to sterile-neutrino production assumptions.

Third, the thermal relic scenario with suppressed temperature provides the weakest cosmological pressure among the sterile-neutrino realizations considered here under the adopted parameterization. In this case, stringent limits remain on the effective sterile abundance and effective sterile mass, but the underlying physical sterile mass can be much less tightly constrained because the relic population is colder and less abundant. This illustrates that cosmological viability is achieved not by allowing a large hot-dark-matter density, but by suppressing the sterile population sufficiently that its gravitational imprint remains small. We note, however, that the DTS and DW scenarios correspond to different prior structures in the derived $\{\Delta N_{\rm eff},m_s^{\rm eff}\}$ parameter space, and therefore part of the quantitative difference between their Bayesian performances reflects the adopted parameterization in addition to the underlying physics.

A key conclusion of this work is therefore that current cosmological data do not generically exclude sterile neutrinos as a broad physical category. Rather, they strongly constrain the combination of sterile abundance and mass, with the most severe pressure applying specifically to scenarios involving full thermalization or large effective abundance. The viability of sterile neutrinos depends critically on the underlying production mechanism, thermal history, and resulting phase-space distribution.

Overall, our results emphasize that future sterile-neutrino studies must move beyond the traditional fully thermalized paradigm. Precision cosmology is increasingly capable of probing the detailed properties of relic particle populations, but robust conclusions require careful consideration of the specific microphysical realization under study. As current and upcoming surveys~\cite{Euclid:2024imf,Lin:2022aro,Nascimento:2025bax,Ballardini:2021frp} further improve constraints on radiation density, large-scale structure, and the late-time expansion history, the distinction between different sterile-neutrino production histories will become increasingly important.
More broadly, the results presented here highlight that cosmological constraints on sterile neutrinos should be interpreted within the context of specific production mechanisms rather than in terms of a single benchmark realization. The future interplay between cosmological observations and laboratory searches will therefore be essential for determining whether a sterile-neutrino sector exists and, if so, how it was populated in the early Universe.

\begin{acknowledgments}
%\bigskip
\noindent We thank the referee for the careful reading of the manuscript and for several constructive comments and suggestions. R.C.N. thanks Sunny Vagnozzi for the valuable discussions regarding the results presented in this work. R.C.N. thanks the financial support from the Conselho Nacional de Desenvolvimento Científico e Tecnológico (CNPq, National Council for Scientific and Technological Development) under the project No. 304306/2022-3, and the Fundação de Amparo à Pesquisa do Estado do RS (FAPERGS, Research Support Foundation of the State of RS) for partial financial support under the project No. 23/2551-0000848-3. 
E.D.V. is supported by a Royal Society Dorothy Hodgkin Research Fellowship. This article is based upon work from the COST Action CA21136 - ``Addressing observational tensions in cosmology with systematics and fundamental physics (CosmoVerse)'', supported by COST - ``European Cooperation in Science and Technology''.
S.G.\ is supported by the Research grant TAsP (Theoretical Astroparticle Physics) funded by Istituto Nazionale di Fisica Nucleare (INFN), through the Ram\'on y Cajal contract RYC2023-044611-I funded by MICIU/AEI/10.13039/501100011033 and FSE+, and by the Spanish grants PID2023-147306NB-I00 and CEX2023-001292-S (MCIU/AEI/10.13039/501100011033).

%\appendix
%\section{Main Tables}

\end{acknowledgments}
\bibliographystyle{apsrev4-1}
\bibliography{main}
\end{document}